\documentclass[a4paper]{article}
\usepackage{amsmath,amstext,amsgen,amsbsy,amsopn,amsfonts,amssymb}
\usepackage{easybmat}
\usepackage{graphics}
\usepackage[pdftex]{graphicx}
\usepackage{epsfig}
\usepackage{amsthm}
\usepackage{bm}
\usepackage{algorithm}
\usepackage{algorithmic}

\usepackage{wrapfig}
\usepackage{hyperref}
\usepackage{arydshln}
\usepackage{verbatim}
\usepackage{fancyvrb}
\usepackage{gensymb}
\begin{document}
\large
\title{\textbf{Correcting Variable Importance Scored by Random Forests}}
\author{
Guancheng Zhou$^{\ddag}$,  Haiping Xu$^{\dag }$, Jason Liu$^{\P }$, Donghui Yan$^{\ddag}$
\vspace{0.1in}\\
%%\vspace{0.1in}\\
$^{\dag}$Computer and Information Science\\[0.04in]
$^\ddag$Mathematics and Data Science\\[0.02in]
University of Massachusetts, Dartmouth, MA\\[0.05in]
$^\P$The Rivers School, Weston, MA%%\\[0.02in]
}

\date{\today}
%%\date{}
\maketitle \normalsize
\begin{abstract}
\noindent
Variable importance produced by Random Forests (RF) is used widely in statistical data analysis, 
and has played an important role in a variety of tasks such as assisting model interpretation, model selection 
and diagnosis, and cost-bounded learning etc. However, the calculation of variable importance in RF does not
take into account of the correlations among variables, and variables that are correlated to many other variables
tend to receive a lower importance index or being completely masked (i.e., with an importance index near zero) 
by other strongly correlated variables. To prevent influence from unwanted correlated variables in calculating
variable importance, we propose to group variables by their conditional correlations (conditional on the response 
variable). We explore two computationally efficient options, with one grouping variables individually, and 
then separates the variable of interest from all correlated variables, while the other uses clustering to group variables 
according to their pairwise conditional correlations. Our experiments show that both lead to sensible 
corrections to the importance of variables.  
\end{abstract}

%%\begin{keywords}
%%Variable selection, ensemble, model schedule, classification
%%\end{keywords}

\section{Introduction}
\label{section:Intro}
Random Forests (RF) \cite{RF} is widely viewed as one of the most powerful tools in statistics and 
machine learning according to many empirical studies \cite{RF,Caruana2006,caruanaKY2008}. 
It is an ensemble of decision trees. 
Starting from the root node (which corresponds to the entire data), each tree is built by a recursive
partition of the tree nodes. At each node, RF randomly samples a number of features (or sets of features) 
and select one that would optimally split that node according to some metric such as increment in
purity as measured by the Gini index \cite{CART, RF, ZhouYan2025c}. This process continues recursively 
until a stopping criterion is met, e.g., fully growing the tree or when a node is pure (i.e., only consists of 
points with the same label for classification). RF achieves a good balance between reductions in bias and 
variance \cite{RF}, thus exhibits excellent performance in many applications. RF is easy to use (e.g., 
very few tuning parameters) and often delivers one of the best empirical performance when compared 
to competing classifiers such as support vector machines (SVM) \cite{SVM}, boosting \cite{AdaBoost, 
Friedman2002,ChenGuestrin2016} etc.  
\\
\\
One nice feature about RF is that it can output the importance of all the variables in the model while performing 
tasks such as classifications. This makes it particularly useful in practice. In many applications, 
especially when there are many variables in use, it is highly desirable to know which variables are important, 
or whether some variables are more important than others. Thus a proper numerical measure of the relative 
importance for all variables is highly desirable. Indeed, variable importance has been widely used in statistical 
data analysis and has played an important role in various tasks such as assisting model interpretation, 
model selection and diagnosis \cite{rsDiagnosis2019}, cost-bounded learning \cite{YanQin2021} etc.  
\\
\\
There are two feature importance measures in RF, with one based on the Gini index \cite{CART,RF} and the other 
on permutation accuracy. The Gini index measures how `mixed' or `pure' of points in a set of points, e.g., a tree node. At 
a given node, $\bm{N}$, let the proportion of data points for class $j$ be denoted by $p_j$ for $j=1,...,J$, then 
the Gini index or impurity of $\bm{N}$ is defined as
\begin{equation*}
Gini(\bm{N})=\sum_{j=1}^J p_j (1-p_j).
\end{equation*}
RF feature importance by the Gini index is calculated by accumulating the total reduction in node impurity (which typically
decreases due to a node split) that a given feature brings to all trees in the forest. It measures how well the given 
feature separates the data, with higher values indicating greater importance. 
\\
\\
%%\\
%%\\
We consider the second feature importance produced by RF here, as it is easier to calculate (i.e., one does not need to track
the entire process of tree growth). Also the Gini importance measures can be biased towards continuous or high cardinality 
(e.g., categorical variables with many more levels) variables \cite{StroblHothorn2007}. The calculation of permutation accuracy 
proceeds as follows. To determine the importance of a given feature, say, the $i\textsuperscript{th}$ feature, one randomly 
permutes values along this feature,  then its association with the response variable $Y$ is broken. When the permuted feature is later 
used along with other features, the predictive accuracy tends to decrease as effectively the permuted feature is destroyed. 
The drop in predictive accuracy can then be used as a measure of the importance of the $i\textsuperscript{th}$ feature. 
\\
\\
While feature importances produced by RF have been widely used in practice, it suffers from a serious problem in the sense that 
correlations among variables are not taken care when calculating the importance of involved features, and the importance of such 
features may be (partially) masked by the correlated features. For example, variable $U_1$ is known to be an important variable 
to the response variable $Y$. However, several variables that are highly correlated with $U_1$ (conditional on $Y$), namely, 
$V_1, V_2, V_3$, are also present in the data. When calculating the importance of $U_1$, one permutes its feature values, and 
then calculates the difference in predictive accuracy due to permutations of feature values on $U_1$. As $V_1, V_2, V_3$ are highly 
correlated with $U_1$, the difference in the resulting predictive accuracy may be quite small, which would imply a small feature 
importance value for variable $U_1$. This is a significant discrepancy from the truth. The main problem is that the calculation of 
permutation feature importance does not account for the correlation among
variables. To alleviate the problem, we propose to first detect (conditionally) correlated variables or group variables by conditional
correlation. Then, removing all the correlated variables when performing permutation feature importance calculation. This will 
effectively prevent correlated variables from playing a role in contributing to the predictive accuracy, thus getting rid of the `annoying' 
influence from correlated variables when calculating the importance of individual variables. 
\\
\\
The remaining of this paper is organized as follows. In Section~\ref{section:method}, we give a detailed description 
of our proposed approach. This is followed by a discussion on related work in Section \ref{section:backgroundRelated}.
In Section~\ref{section:experiments}, we present 
experimental results on some real datasets. Finally, we conclude in Section~\ref{section:conclusions}. 
%%
%%
%%
%%
%%\\
%%\\
%%
%%
\section{The method}
\label{section:method}
To correct variable importances by RF, our main idea is to find ways to get rid of influences from correlated variables 
when computing the importance of a given variable. To achieve this, we try to group variables that are correlated, and 
we explore two different options in grouping the variables. One is to find variables that are correlated to the given 
variable, and the other is to divide the entire set of variables into groups by clustering according to some similarity metric. 
Both options hinge on an important concept, {\it conditional correlation}, which we will describe in the following. 
\\
\\
The {\it conditional correlation coefficient} \cite{KendallStuart1979} between two variables $U$ and $V$, given $Y$, is 
defined as the ratio of their conditional correlation standardized by the product of their conditional standard deviations
\begin{equation*}
\rho(U,V|Y)= \frac{Cov(U,V|Y)}{\sqrt{Var(U|Y) \cdot Var(V|Y)}}
\end{equation*}
where $Cov(U, V|Y)$ is the correlation of $U$ and $V$ calculated over the distribution where $Y$ is fixed, $Var(U|Y)$
is the variance of $X$ given $Y$. The pairwise conditional correlation matrix $\bm{Corr}$ is formed by calculating the 
conditional correlation (given $Y$) of all pairs of variables $V_i$ and $V_j$ at the $(i,j)-th$ position of the matrix.
\\
\\
We will use the conditional correlation to induce similarity between pairs of variables. The pairwise similarity matrix can 
be used to find a grouping among variables by spectral clustering. The variable importance will then be calculated by 
accounting for the grouping (or conditional correlations) among variables.  
%%\\
%%\\
%%\\
%%\\
%%\\
%%\\
%%
%%
%%
%%
%%
%%
%%
%%
%%
%%
\subsection{Algorithmic description}
\label{section:methodAlg}
In this section, we will describe algorithms to implement the two options we adopt to group variables. The first algorithm works 
on each individual 
variables. Let $\bm{V}_{all}$ be the set of all dependent variables. For a given variable $V_i$, the algorithm first detects the 
list of variables conditionally correlated with $V_i$, which is denoted by $\bm{V}_{cor}$. It is known that none of the variables contained 
in the set $\bm{V}_{nc}=\bm{V}_{all} \setminus \bm{V}_{cor} \setminus \{V_i\}$ are conditionally correlated with variable $V_i$. Thus, 
by adding variable $V_i$ back to $\bm{V}_{nc}$, the increase in predictive accuracy by invoking RF on such variables will be 
due entirely to the addition of variable $V_i$. The net increase in predictive accuracy can then be used as the importance of 
variable $V_i$. This is described as Algorithm~\ref{algorithm:corrVI-Individual} (termed as Method 1).
\\
%%\\
%%
%%
%%
%%
\begin{algorithm}
\caption{\it~~corrVI-Individual(X, Y)}
\label{algorithm:corrVI-Individual}
\begin{algorithmic}[1]
\FOR {$i=1$ to $p$}
	\STATE Compute $\bm{V}_{cor}$ the set of variables conditionally correlated with $V_i$ using inout data $(X,Y)$; %%
	\STATE Invoke RF on variable sets $\bm{V}_{all} \setminus \bm{V}_{cor}$ and $\bm{V}_{all} \setminus \bm{V}_{cor} \setminus \{V_i\}$, respectively; %%
		\STATE Let the difference in the two respective predictive accuracies be $\alpha_{\delta}$; %%
		\STATE Report $\alpha_{\delta}$ as the importance of variable $V_i$; %%
\ENDFOR%%
\RETURN(importance of all variables); %%
\end{algorithmic}
\end{algorithm} 
\\
In Algorithm~\ref{algorithm:corrVI-Individual} and more generally in our approach, we seek to calculate the reduction in predictive accuracy 
after removing variables that are conditionally correlated to a given variable $V_i$. In this process, we are actually removing `more' information 
than needed in the sense that $\bm{V}_{cor}$ contains predictive information beyond that entailed by variable $V_i$. The calculated reduction 
in predictive accuracy from variable sets $\bm{V}_{nc} \cup \{V_i\}$ to $\bm{V}_{nc}$ will be slightly overestimated than in the ideal case where 
no information beyond  $V_i$ is dropped (c.f. discussion in Section 3.2.1 in  \cite{TACOMA} where the amount of separation between different 
classes reduces due to loss of information). However, if the predictive accuracy on variables $\bm{V}_{nc} $ is `decent' (i.e., a reasonably big 
separation between different classes), then the amount of overestimation would be negligible. 
\\
\\
The second algorithm works differently in that it first divides the entire set of variables $\bm{V}_{all}$ into disjoint subgroups of variables,
$\bm{V}_{all} = \bm{G}_1 \cup \cdots \cup \bm{G}_K$, such that variables from different subgroups are not or only weakly correlated. This 
is done by treating the 
pairwise conditional correlation matrix as a pairwise similarity matrix thus spectral clustering can be readily applied to split $\bm{V}_{all}$ 
into subgroups. To assess the importance a variable, say $V \in \bm{G}_i$, one can simply remove the entire set of variables $\bm{G}_i$ 
from $\bm{V}_{all}$, and adding back variables $V$ and declare the increase in predictive accuracy by RF as the importance of variable $V$.
This is sensible as no variable in set $\bm{V}_{all} \setminus \bm{G}_i$ is conditionally correlated with variable $V$ (or conditional 
correlation can be neglected). Thus the importance of variable $V$ will not be masked by any variables in the set $\bm{V}_{all} \setminus \bm{G}_i$.
So the difference in predictive accuracy would be caused entirely by adding variable $V$, and this difference can be used as the importance
of variable $V$. The computing procedure is described as Algorithm~\ref{algorithm:corrVI-M2} (termed as Method 2).
\begin{algorithm}
\caption{\it~~corrVI-Spectral(X, Y)}
\label{algorithm:corrVI-M2}
\begin{algorithmic}[1]
\STATE Compute pairwise conditional correlation matrix $\bm{Corr}$ over all variables; %%
\STATE Computer similarity matrix $\bm{W}=\exp(\bm{Corr}/\sigma^2)$ for suitable value of $\sigma$;%% 
\STATE Apply spectral clustering to $\bm{W}$ to divide $\mathcal{V}_{all}$ into subgroups, $\bm{G}_1, ..., \bm{G}_K$; %%
\FOR {$i=1$ to $K$}
	\STATE Let $\bm{V}_{nc}=\bm{V}_{all} \setminus \bm{G}_i$; %%
	\FOR {Each $V \in \bm{G}_i$}
		\STATE Let $\bm{V}_r=\bm{V}_{nc} \cup \{V\}$; %%
		\STATE Invoke RF to get predictive accuracy on set of variables $\bm{V}_{nc}$ and $\bm{V}_r$; %%
		\STATE Let the difference in the two respsective predictive accuracies be $\alpha_{\delta}$; %%
		\STATE Report $\alpha_{\delta}$ as the importance of variable $V$; %%
	\ENDFOR%%
\ENDFOR
\RETURN(importance of all variables); %%
\end{algorithmic}
\end{algorithm} 
%%
%%
%%
%%
%%\\
%%\\
%%\noindent
\subsection{A working example}
\label{section:toyExp}
\begin{figure}[htb] 
\begin{center}
\includegraphics[scale=0.3,clip, angle=0]{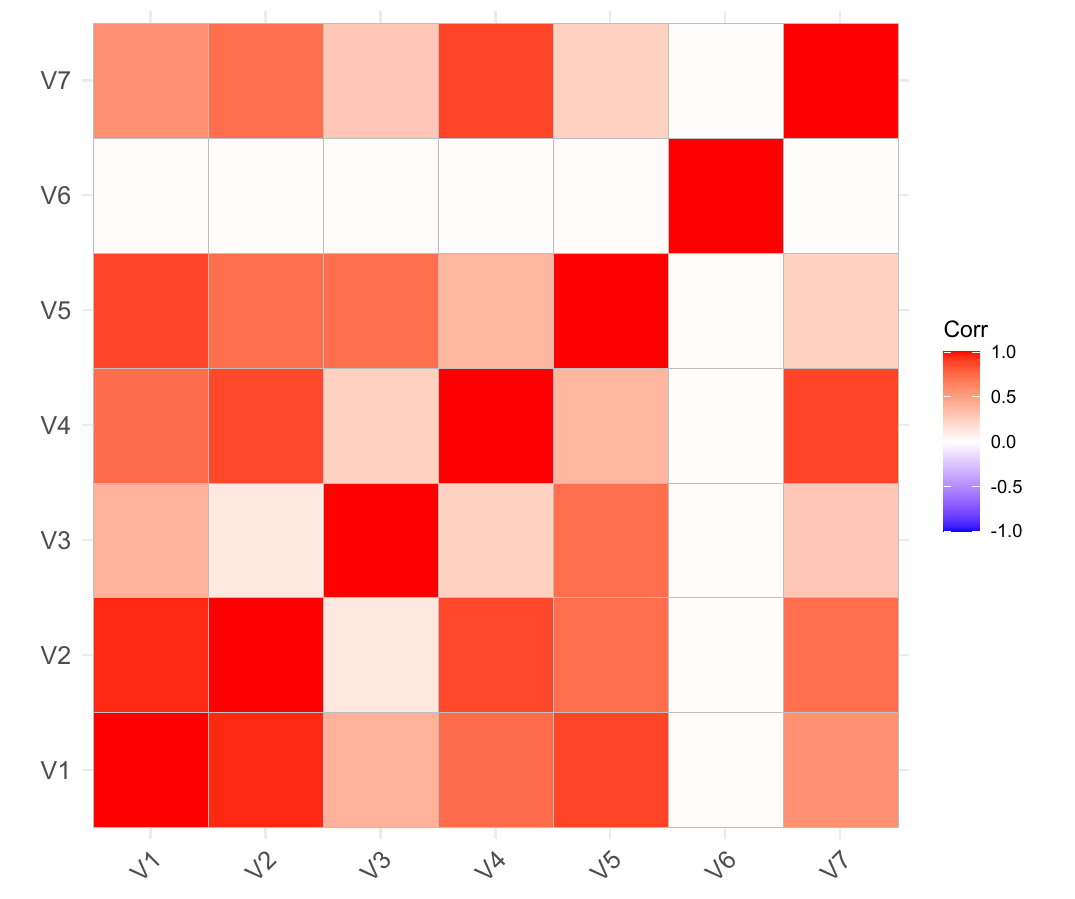}\hspace{0.01in}
\raisebox{-0.1in}{
\includegraphics[scale=0.35,clip, angle=0]{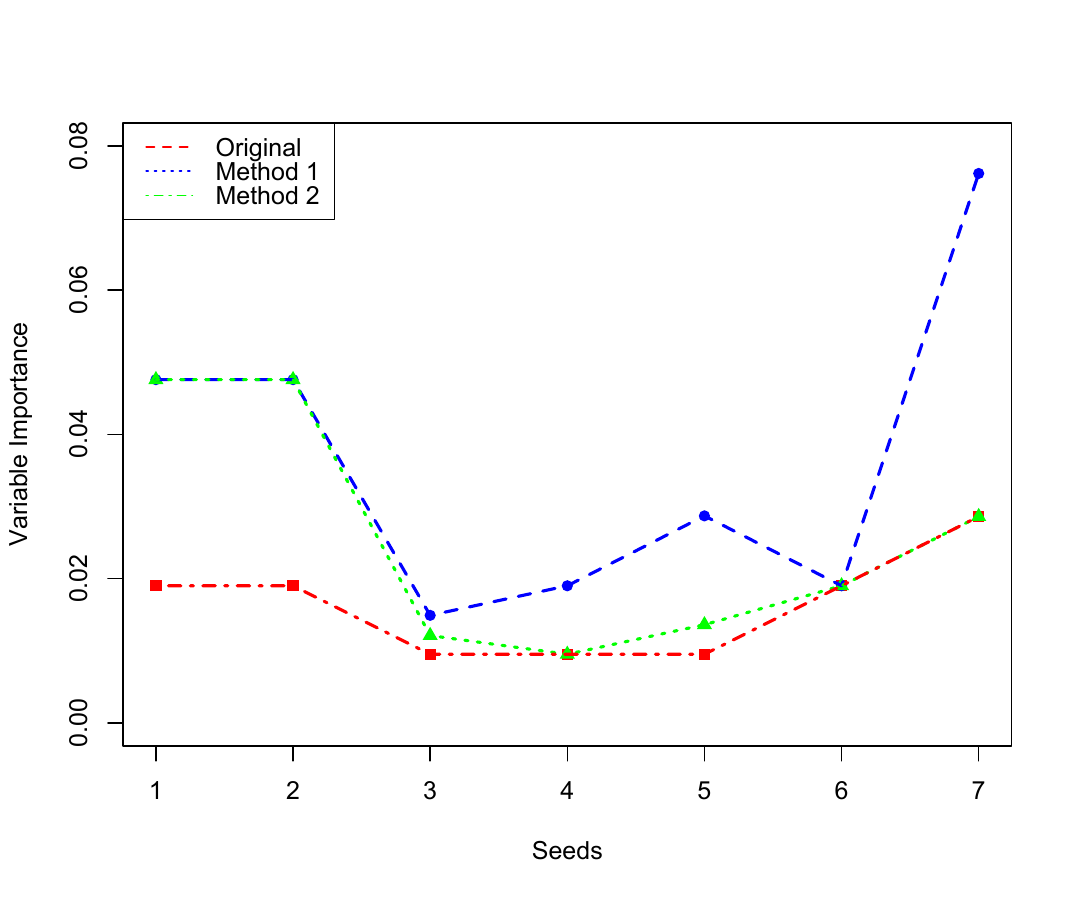}}
%%\abovecaptionskip=-1pt
\caption{\it The pairwise conditional correlation matrix of all variables in the UC Irvine Seeds dataset, and the importance
value given by the original RF, Method 1 and Method 2, respectively.}
\label{figure:gramSeeds}
\end{center}
\end{figure}

We will use the Seeds dataset from UC Irvine Machine Learning Repository \cite{UCI} as a working example to illustrate 
the idea of our algorithm. There 
are 7 variables in the data: 1) area A ($V_1$); ~2) perimeter P ($V_2$); ~3) compactness $C = 4*pi*A/P^2$ ($V_3$); 
~4) length of kernel ($V_4$); ~5) width of kernel ($V_5$);  ~6) asymmetry coefficient ($V_6$); ~7) length of kernel groove ($V_7$). 
All are measured geometric parameters of wheat kernels. The response variable is the wheat seeds types, which takes value of 
one of Kama, Rosa, or Canadian. The conditional correlational matrix of all variables 
is visualized as the left panel of Figure~\ref{figure:gramSeeds}.
\\
\\
In Method 1, first we compute, for each variable, the set of conditionally correlated variables, and this is summarized as 
Table~\ref{table:cvvSeeds}. Note that $V_6$ is not conditionally correlated any other variable, so its conditionally
correlated set is empty and we omit it from the table. 
\begin{table}[h]
\begin{center}
\begin{tabular}{c|c|c|c|c|c|c}
\hline
 \bf{Variable} &  $V_1$           & $V_2$     	& $V_3$   	&  $V_4$    & $V_5$        	&$V_7$\\
\hline
\bf{Correlated variables} &$V_2$ 		&$V_1$		&$V_1, V_5$	&$V_1, V_2, V_7$   &$V_1, V_2, V_3$	&$V_1, V_2, V_4$\\
\hline
\end{tabular}
\caption{\it Variables in the UC Irvine Seeds dataset and the conditionally correlated variables for each variable. $V_6$ is omitted
from the table as it is not conditionally correlated to other variables.} 
\label{table:cvvSeeds}
%%\end{minipage}
\end{center}
\vskip -0.1in
\end{table}
Then, for each variable, we carry out procedures as in Algorithm~\ref{algorithm:corrVI-Individual}, and obtain the importance
for each variable. The variable importances are shown as the blue dashed line in the right panel of Figure~\ref{figure:gramSeeds}. 
Note that, for each variable, the conditionally correlated variables are ranked by their correlation coefficients, and then a sequential 
check is performed and, if a variable does not noticeably change the classification accuracy then it will not show up in the table. 
\\
\\
It can be seen that the importance of variables $V_1, V_2, V_7$ are substantially adjusted upwards while that for variables 
$V_3, V_4, V_5$ are only slightly adjusted upwards. This is expected, as we know from knowledge about wheat seeds that 
the area ($V_1$), perimeter ($V_2$) and length of the kernel groove ($V_7$) are all important in distinguishing the wheat seeds
types. The length of the kernel groove ($V_7$) is a primary morphological feature used to classify and identify seed types, 
so we expect it has the largest importance value. There is a strong positive correlation (0.92) between the overall length 
of a kernel ($V_4$) and the length of its groove, making it a reliable parameter for characterizing wheat seed size and shape 
thus an importance value as adjusted is reasonable. In the original importance obtained by RF, the importance of compactness 
($V_3$) is nearly zero. As the compactness is a function of the area and the perimeter, and given that the later two are present, 
the role of compactness would be redundant thus a near 0 importance. But if one looks at the compactness alone, this variable 
does contain information that helps distinguish the wheat seeds, so its importance should not be zero, and we think the correction 
is needed and a small importance value would be reasonable. Indeed its adjusted importance value, which is much smaller than that of 
$V_1$ and $V_2$, is expected as $V_3$ is a function of $V_1$ and $V_2$ thus diluting the distinguishing power of each of $V_1$
and $V_2$. Moreover, by intuition, the length and width of the kernel should have similar distinguishing power for the wheat seeds 
types, thus a similar variable importance value will be more appropriate, as corrected by Method 1.   
\\
\\
In Method 2, first we calculate the pairwise conditional correlations for all the 7 variables in the data (c.f., left panel of 
Figure~\ref{figure:gramSeeds}). 
Then we will use spectral clustering to find a grouping among the 7 variables in the Seeds dataset. Treating the pairwise 
conditional correlations matrix as the similarity matrix, and introduce the following kernel
\begin{equation*}
K(u,v) = e^{corr(u,v)/\sigma^2},
\end{equation*}
where $u$ and $v$ are the two variables involved, and $corr(,)$ stands for the conditional correlation. Applying spectral 
clustering with a proper choice of $\sigma$ leads to the following grouping of the variables in the Seeds dataset:
~$\{V_1,V_2\}$, ~$\{V_3,V_5\}$, ~$\{V_6\}$, ~$\{V_4,V_7\}$.
Applying our variable importance correction algorithm Method 2 leads to importance values (marked by green dashed line)
in the right panel of Figure~\ref{figure:gramSeeds}.
\\
\\
Method 2 similarly adjusts the importance of variables $V_1, V_2$ upwards significantly, this is consistent with our knowledge 
about the wheat seeds as well as Method 1. However, the importance of variables $V_3, V_4, V_5$ nearly do not receive 
adjustment compared to that by Method 1. This implies that the compactness, and the length and width of the kernel do not 
play much role in distinguishing wheat seeds types. This deviates a little from the truth, and we attribute this to the quality of 
the clustering (the cluster between variables $V_4$ and $V_5$ is not captured properly, due to the much stronger affinity between 
variables $V_4$ and $V_7$). In comparison, Method 1 does much better about this (because of its flexibility in the 
sense that correlated variables do not necessarily form disjoint clusters). The biggest difference lies in variable $V_7$,  which 
does not receive any adjustment from Method 2 but a major upwards adjustment by Method 1. Again we attribute this to the 
fact that the strong affinity between variable $V_7$ and $V_1, V_2$ is not captured by grouping via spectral clustering (though 
its correlation with variable $V_4$ is taken care). As a result, the contribution by variable $V_7$ is partially masked by variable 
$V_1$ and $V_2$. In comparison, Method 1 is more flexible that Method 2, as it is not always true that those correlated variables 
must form disjoint subgroups. 
\\
\\
It is interesting to note that, for variable $V_6$, its importance under all three approaches remains the same. We attribute this 
to the fact that variable $V_6$ is nearly not correlated to all other variables (as evident by the entire white row/column in the 
pairwise conditional correlation matrix or the omission of $V_6$ from Table~\ref{table:cvvSeeds}), thus no masking phenomenon 
among correlated variables is present (i.e., our correction does not affect the importance of variable $V_6$). 
%%\\
%%\\
%%
%%
\subsection{Removal instead of permutation of features}
\label{section:featureRemoval}
\begin{figure}[h] 
\begin{center}
\includegraphics[scale=0.52,clip, angle=-90]{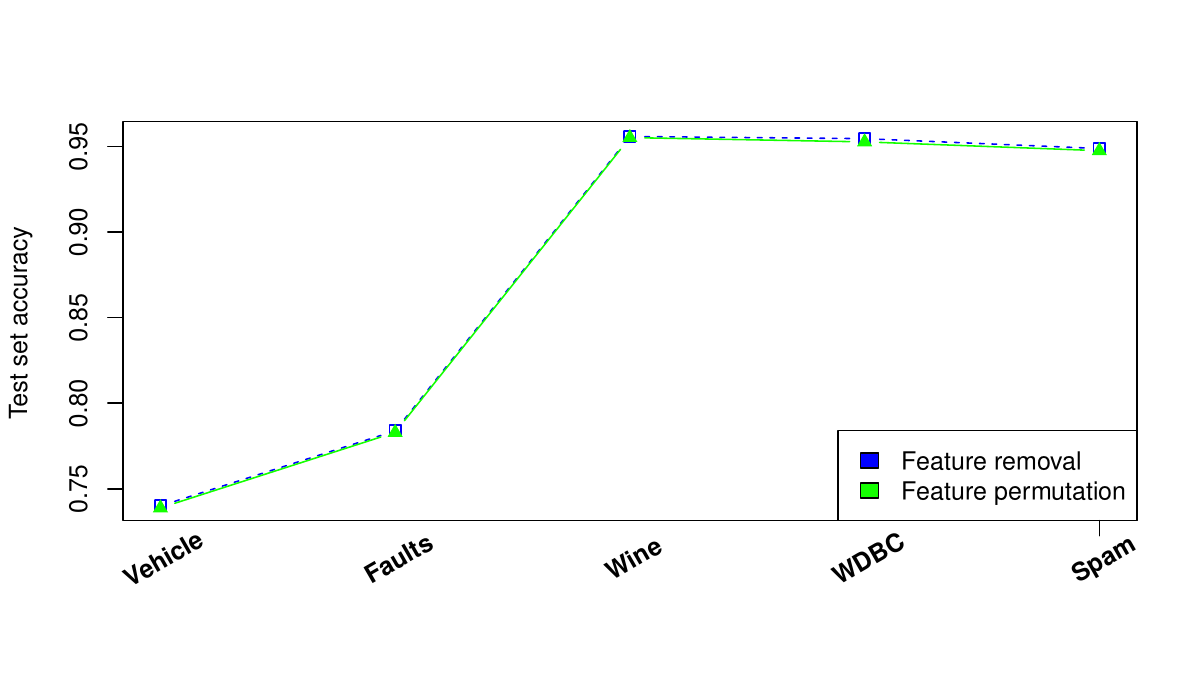}
\abovecaptionskip=-3pt
\caption{\it Test set accuracy of some UC Irvine datasets when removing or permuting a variable. }
\label{figure:removalVSPermu}
\end{center}
\end{figure}

\begin{figure}[htb] 
\begin{center}
\includegraphics[scale=0.52,clip, angle=-90]{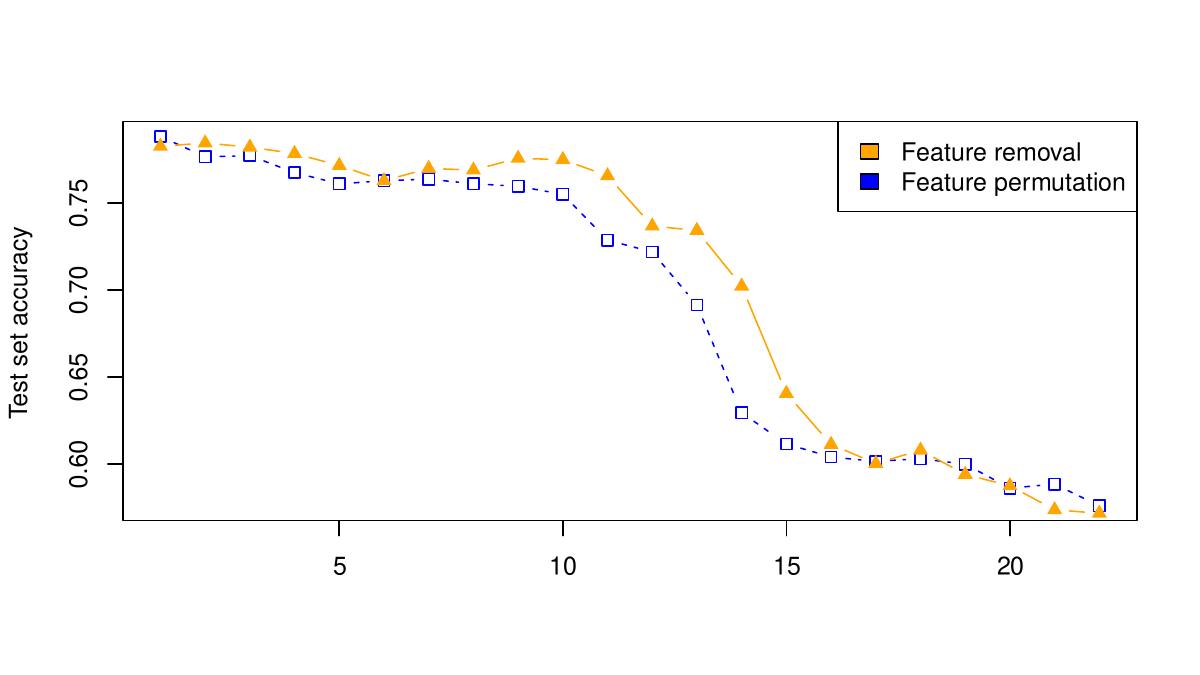}
\abovecaptionskip=-3pt
\caption{\it Test set accuracy of the UC Irvine Concrete Faults dataset when sequentially removing or permuting its variables. Shown in 
the x-axis are variable indices. }
\label{figure:removalsVSPermus}
\end{center}
\end{figure}

One may notice that in our variable importance correction algorithms, the variables involved are removed instead of permuted. This is
done for the following considerations. First of all, we note that the difference caused in the resulting predictive accuracy is negligible by 
removing or permuting a single variable. This is due to the strong built-in variable selection capability of RF. As a variable is permuted, 
its association with the response variable is broken, thus this variable becomes essentially a noise variable, and at each node split, the 
noise variable has almost a negligible chance to be selected as the splitting variable. So permutation achieves the same effect as 
removing the variable. This is demonstrated in Figure~\ref{figure:removalVSPermu} where the resulting predictive accuracy under 
permutation or removal of a variable nearly overlap. 
\\
\\
However, the behavior in predictive accuracy becomes different when permuting multiple variables. Imagine there are multiple pure 
noise variables, then the probability that a noise variable is being selected at a node split increases (actually occurs when sampled 
candidate variables are all noise variables). This will cause the predictive accuracy to decrease more than simply removing variables 
(s.t. the non-noise variable being the same). As shown in Figure~\ref{figure:removalsVSPermus}, the performance gap is no longer 
negligible when permuting three or more, and starts deteriorating when permuting more variables. Thus, permuting variables is no 
longer a proper way in getting rid of the influence of several variables, and instead we will simply remove those conditionally correlated 
variables in correcting the importance of variables.   
\section{Background and related work}
\label{section:backgroundRelated}
In this section, we discuss necessary background and work related to ours. We start by an introduction to spectral clustering as 
it is used by Method 2.
%%\\
%%\\
\subsection{Spectral clustering}
\label{section:introSPC}
\begin{figure}[h] 
  \begin{center}
    \includegraphics[width=0.45\textwidth]{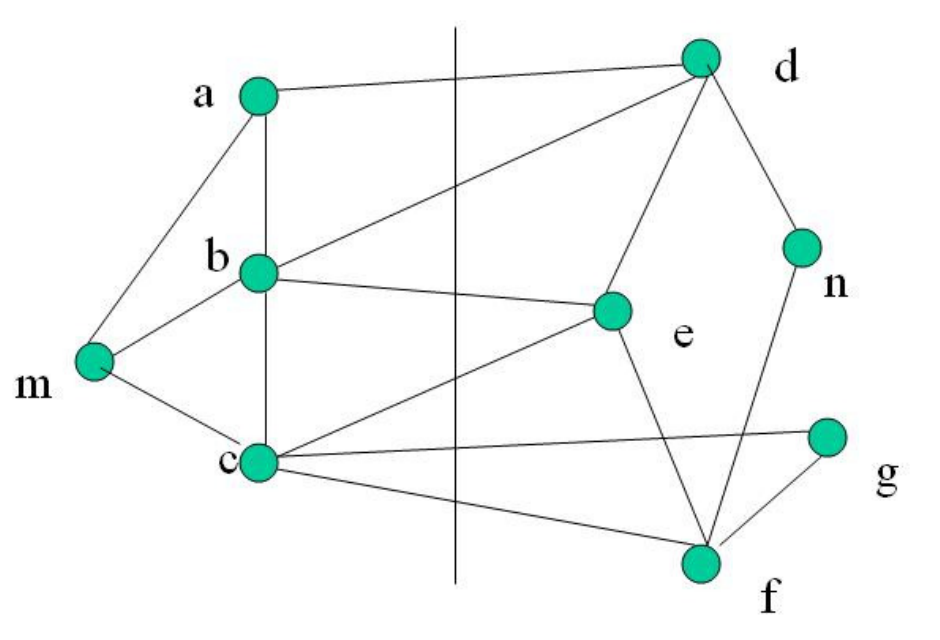}%{./Pictures/mainscreen1.png}
    \caption{\it Affinity graph and graph cut. The cut is given by the set $\{ad,bd,be,ce,cg,cf\}$, which partitions the 
    vertices of the graph into 
    $V=V_1 \cup V_2=\{a,b,c,m\} \bigcup \{d,e,f,n,g\}$.}
    \label{figure:graphCut}
  \end{center}
\end{figure} 
%%\\
%%\\
%%\\
\noindent
Spectral clustering is used in Method 2 to group conditionally correlated variables, with which one can evaluating 
the importance of a variable without being influenced by conditionally correlated variables. It is a class of clustering 
algorithms that works on the pairwise similarity between points, instead of directly on the distance between points. It 
works on an affinity graph over data points $X_1,...,X_N$ and seeks to find a ``minimal" graph cut. 
Depending on the choice of the similarity metric and the objective function to optimize, there are a number of variants 
for spectral clustering, including \cite{ShiMalik2000, NgJordan2002, YanHuangJordan2009}. Our discussion will be 
based on normalized cuts of Shi and Malik \cite{ShiMalik2000}. It arises from the image segmentation problem in computer
vision. An {\it affinity graph} is defined as a weighted graph $\mathcal{G} = (V, \mathcal{E}, A)$
where
\begin{eqnarray*}
&&V = \{X_1,...,X_N\}, ~A = (a_{ij})_{i,j=1}^N, \\
&& \mathcal{E} = \{(i,j):~ i, j=1,...,N, i \neq j\}
\end{eqnarray*}
with $a_{ij}$, the weight of edge $(i,j) \in \mathcal{E}$, encoding the affinity (or similarity) of data points $X_i$ and $X_j$. 
The matrix $A$ is often called the similarity matrix or affinity matrix. Figure~\ref{figure:graphCut}
illustrates affinity graph and a graph cut. 
\\
\\
Let $V = (V_1, \ldots, V_K)$ be a partition of $V$. Define
\begin{equation*}
W(V_1, V_2) = \sum_{i \in V_1, j \in V_2} a_{ij}
~\mbox{for}~V_1, V_2 \subseteq V.
\end{equation*}
Normalized cuts seeks to solve the following optimization problem
\begin{equation*}  \arg\min_{V_1,...,V_K \subseteq V} \sum_{j=1}^K \frac{W(V_j, V) -
W(V_j, V_j)} {W(V_j, V)}.
\end{equation*}
The above is an integer programming problem thus intractable, a relaxation to real values leads 
to an eigenvalue problem for the Laplacian matrix.
\begin{equation}
\label{eq:defLaplacian}
\mathcal{L}_A=D^{-\frac{1}{2}}(D-A)D^{-\frac{1}{2}},
\end{equation}
where $D=diag(d_{1},...,d_{N})$ is the degree matrix with $d_i=\sum_{j=1}^N a_{ij},i=1,...,N$. 
Normalized cuts look for the second smallest eigenvector of $\mathcal{L}_A$, and round its 
components to produce a bipartition of the graph. Similar procedure can be applied to each 
of the partitions recursively until a stopping criterion is met. 
%%
%%
%%\\
%%\\
%%
%%
\subsection{Related work}
\label{section:related}
One line of work that is closely related to our is \cite{StroblHothorn2007}. The difference is that \cite{StroblHothorn2007} 
works on correcting biases towards continuous or high cardinality (e.g., categorical variables with many more levels) 
variables while we correct biases in variable importance due to multi-collineariy. Another work that is closely related is 
SHAP \cite{SHAP2017} where the importance of a given variable $V$ is 
computed as the difference in predictive accuracy over all possible subsets not including $V$ versus after the variable $V$ 
is added, and then weighted averaged by their respective contributions. SHAP possesses several nice features but the 
computational challenge for high dimensional data is prohibitive. Also related is DeepLIFT \cite{DeepLIFT2017} which 
decomposes the prediction of a neural network on a specific input by back-propagating the contributions of all neurons 
in the network to features at the input. DeepLIFT compares the activation of each neuron to its `reference activation' 
and assigns contribution scores according to the difference. However, DeepLIFT applies only to neural network algorithms.
\\
\\
Additionally, a popular work, LIME \cite{LIME2016}, is related to ours through the lens of model interpretation. It attempts 
to interpret the model behavior by approximating the model locally at the neighborhood of a given instance. Then a simple
sparse linear model is trained on the perturbed data (perturbed around the given instance), giving higher weight to instances 
closer to the original instance. The coefficients of the simple linear model will represent the contribution of each feature to 
the specific prediction.
%%
%%
%%
%%\\
%%\\
%%
%%
%%
%%
%%
%%
%%
%%
%%
%%
%%
\section{Experiments}
\label{section:experiments}
We conduct experiments on eight datasets from the UC Irvine Machine Learning Repository \cite{UCI}. The
UC Irvine datasets are the 
{\it Bank marketing, Indian liver patient, Seeds, Wine, Cleveland hearts, Maternal health risk}, and {\it Obesity levels}. 
For the Bank marketing data, we follow \cite{distStatArXiv2019} where all features are converted into numeric values 
and then standardize. A summary of the datasets used is given in Table~\ref{table:datasets}. For the rest of this section, 
we will report variable importance corrections by our approaches for each individual datasets. 
%%\\
%%\\
%%\\
%%\\
%%\\
%%\\
\begin{table}[htb]
\begin{center}
%%\setlength{\extrarowheight}{1pt}
%%\begin{small}
%%\begin{minipage}[b]{0.5\linewidth}%%\centering
\begin{tabular}{r|rrr}
\hline
   \bf{Dataset}                    				& \bf{Features}     	& \bf{Classes}   	&  \bf{Instances}\\[1pt]
\hline \\[-10pt]
Bank marketing						& 16					&2					&45211\\[1pt]
%%Australian	credit approval		&14		&2			&690\\
Indian liver patient			&10		&2			&579\\
Seeds					&7		&3			&210\\
Wine						&13		&3			&178\\
Heart					&13		&5			&303\\
Maternal health risk			&6		&3			&1013\\
Obesity levels				&16		&7			&2111\\
\hline
\end{tabular}
\caption{\it A summary of datasets used in experiments.} 
\label{table:datasets}
%%\end{minipage}
\end{center}
%%\vskip -0.2in
\end{table}
%%%%%%%%%%%%%%%%%%%%%%%%%%%%%%%%%%%%%%%%%%%%%%%%%%%%%%%%
%%%%%%%%%%%%%%%%%%%%%%%%%%%%%%%%%%%%%%%%%%%%%%%%%%%%%%%%
%%
%%
%%\\
%%\\%%[-0.05in]
%%\\
%%\\
%%\\
\subsection{Indian liver patients dataset}
\label{data:IndianLiver}
\begin{figure}[htb] 
\begin{center}
\includegraphics[scale=0.3,clip, angle=0]{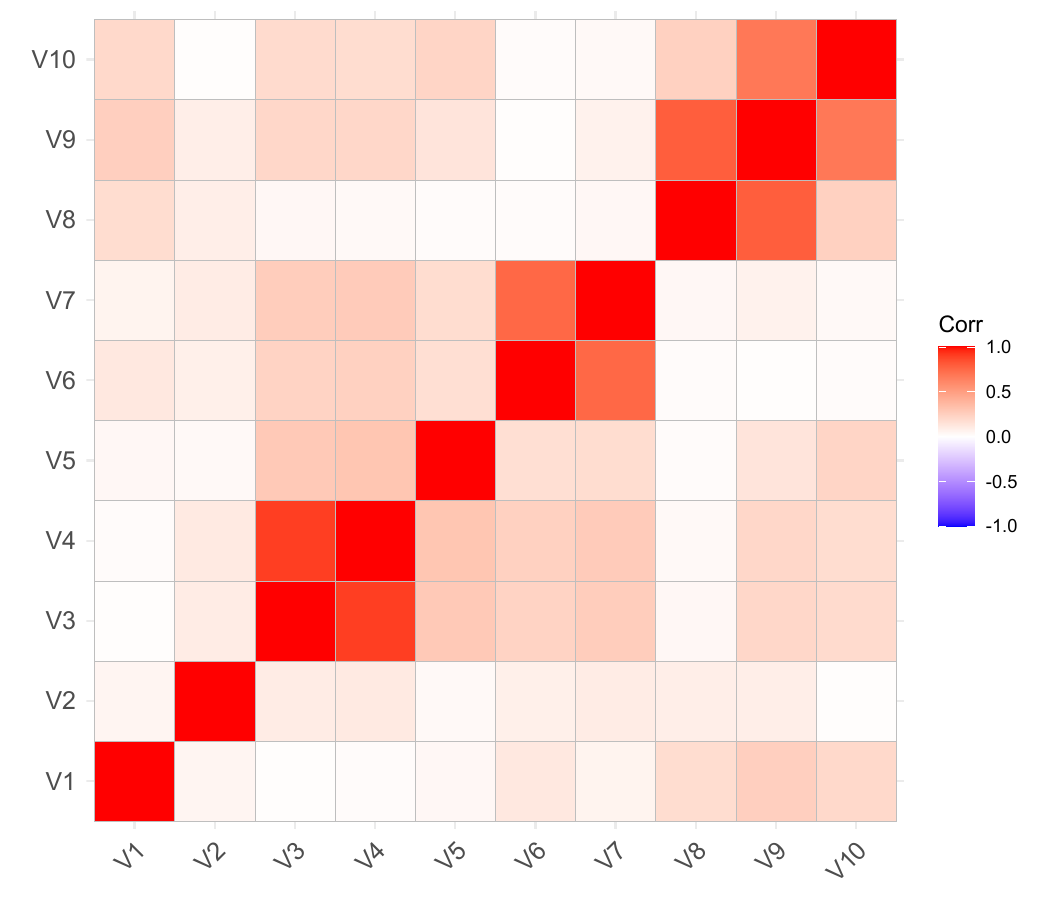}\hspace{0.01in}
\raisebox{-0.1in}{
\includegraphics[scale=0.35,clip, angle=0]{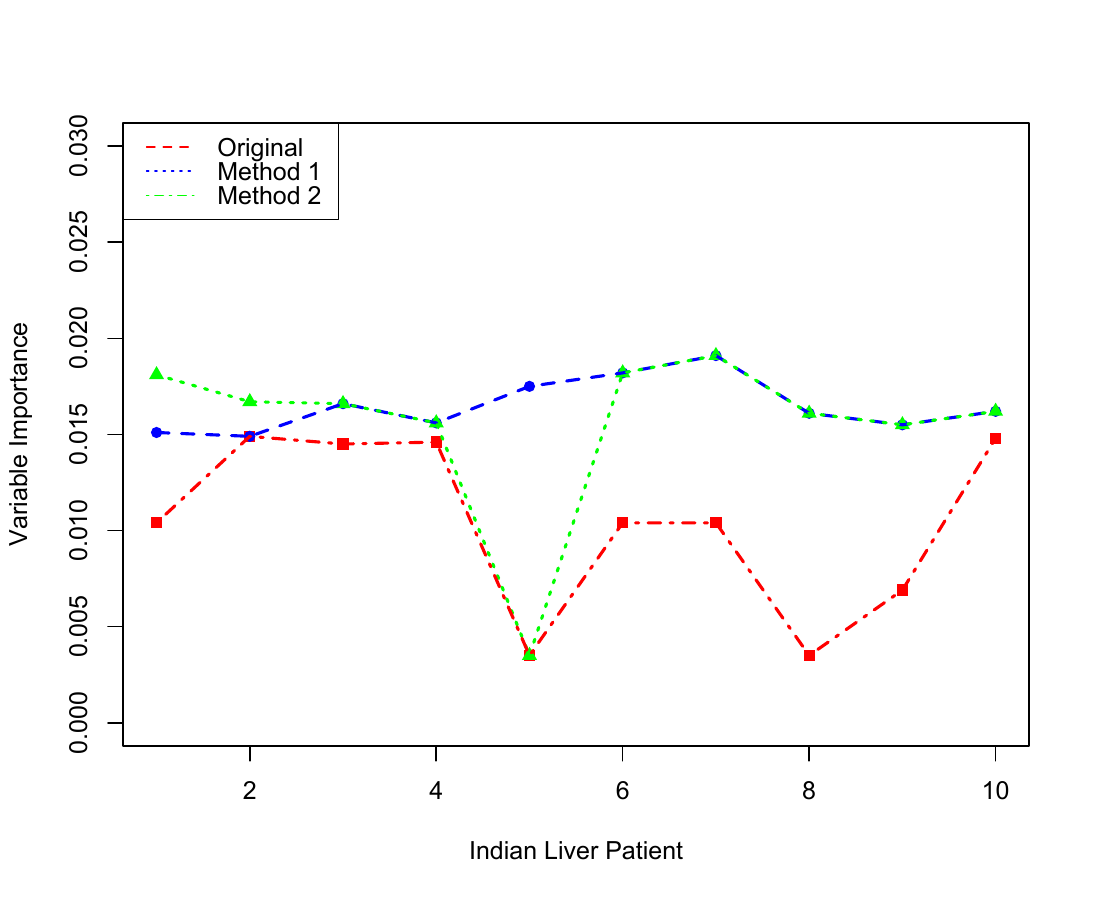}}
%%\abovecaptionskip=-1pt
\caption{\it The pairwise conditional correlation matrix (left panel) of all variables in the UC Irvine Indian liver patients dataset, 
and the importance value (right panel) given by the original RF, Method 1 and Method 2, indicated by red, blue and green 
dashed lines, respectively.}
\label{figure:gramIndianLiver}
\end{center}
\end{figure} 
%%\\
%%\\
There are 10 variables in the UC Irvine Indian liver patients dataset, and are listed as follows: 1) Age ($V_1$); 
~2) Gender  ($V_2$); ~3) total Bilirubin  ($V_3$); ~4) direct Bilirubin  ($V_4$); ~5) Alkphos  ($V_5$); ~6) SGPT  ($V_6$); 
~7) SGOT  ($V_7$); ~8) total proteins  ($V_8$); ~9) albumin  ($V_9$); ~10) A/G ratio  ($V_{10}$). 
Method 1 leads to the set of conditionally correlated variables for each
variable in the dataset as shown as Table~\ref{table:ccvIndianLiver}. Note that henceforth we adopt the following 
conventions: 1) If a variable is missing in the list of variables of interest, then that means there are no other variables 
conditionally correlated with this variable; 2) If multiple variables are shown as variables of interest, that means they are
from a group where any one variable from the group is conditionally correlated with all others.
\begin{table}[H]
\begin{center}
\begin{tabular}{r|l}
\hline
   \bf{Variable}                    				& \bf{Conditionally correlated variables}    \\
\hline 
$V_1$		&$V_9$\\
%%$V_2$			&		\\
$V_3, V_4$			&$V_3, V_4$		\\
$V_5$			&$V_4$		\\
$V_6, V_7$			&$V_6, V_7$		\\
$V_8, V_9, V_{10}$			&$V_8, V_9, V_{10}$		\\
\hline
\end{tabular}
\caption{\it Variables and their conditionally correlated variables for the UC Irvine Indian Liver Patients dataset.} 
\label{table:ccvIndianLiver}
%%\end{minipage}
\end{center}
%%\vskip -0.2in
\end{table}
\noindent
The most notable corrections by Method 1 are to variables $V_5$ and $V_8$. Alkaline Phosphatase (AlkPhos) is 
a critical marker in liver treatment. It is an enzyme found throughout the body, but high levels often signal that liver 
cells lining the bile ducts are damaged or damaged or the bile flow. Total protein levels are crucial in liver treatment 
and diagnosis. They measure albumin and globulins, reflecting the liver’s ability to synthesize proteins, assess nutritional 
status, and detect damage from cirrhosis or hepatitis. Low total protein often indicates decreased liver function, malnutrition, 
or poor absorption. However, the original RF produced nearly 0 importance for these two variables which is clearly 
not appropriate. 
\\
\\
Similarly, $V_6$ acts as a primary biomarker for detecting liver inflammation, damage, or disease. Elevated levels 
often alert doctors to conditions like hepatitis, fatty liver, or drug-induced damage, and $V_7$ is an important diagnostic 
marker in liver treatment and monitoring. Elevated levels indicate damage to liver cells, aiding in the diagnosis of 
conditions like hepatitis, fatty liver, and cirrhosis. It is often used alongside $V_6$ to assess liver damage severity. 
$V_9$ is crucial in treating advanced liver disease to manage complications. It acts as a plasma expander, regulating 
fluid balance, treating ascites, and helping prevent kidney failure in patients with spontaneous bacterial peritonitis (SBP) 
or after paracentesis. So the importance of $V_6, V_7, V_9$ should also be adjusted upwards to reflect their medical
significance. 
\\
\\
For Method 2, applying spectral clustering with a proper choice of $\sigma$ leads to the following grouping of variables 
in the Indian liver patients dataset: ~$\{V_1,V_2\}$, ~$\{V_3,V_4\}$, ~$\{V_5\}$, ~$\{V_6, V_7\}$, ~$\{V_8,V_9, V_{10}\}$.
The resulting importance for most of the variables by Method 2 are similar to that by Method 1. The main difference is on
$V_5$ which has importance of nearly 0 under Method 2. As argued above, $V_5$ appears to be one of the most important
variables, so correction by Method 1 makes sense while Method 2 is not reasonable in not adjusting the importance of $V_5$. 
Algorithmically, Method 2 fails to cover the correlation between $V_5$ and $V_4$ thus the reported importance of $V_5$ is 
partially masked by $V_4$.  
\\
\\
As all the variables in the Indian liver patients dataset were carefully chosen according to years' medical research, most of 
them should be relevant, and indeed variable importances reported by Method 1 and Method 2 (excluding $V_5$) is consistent 
to this observation. 
\subsection{Hearts dataset}
\label{data:Hearts}
There are 13 variables in the UC Irvine Hearts dataset, and are listed as follows: 1) age ($V_1$); ~2) gender  ($V_2$); 
~3) chest pain type  ($V_3$); ~4) resting blood pressure ($V_4$); ~5) serum cholesterol ($V_5$); 
~6) fasting blood sugar ($V_6$); ~7) resting electrocardiographic results ($V_7$); 
~8) maximum heart rate achieved ($V_8$); ~9) exercise induced angina  ($V_9$); 
~10) oldpeak  (ST depression induced by exercise relative to rest, $V_{10}$); 
~11) slope of the peak exercise ST segment ($V_{11}$); 
~12) ca: number of major vessels (0-3) colored by flourosopy  ($V_{12}$); 
~13) thallium heart scan on potential defects of blood flow to heart muscle ($V_{13}$).
The response variable is diagnosis of heart disease (angiographic disease status).
\begin{figure}[htb] 
\begin{center}
\includegraphics[scale=0.28,clip, angle=0]{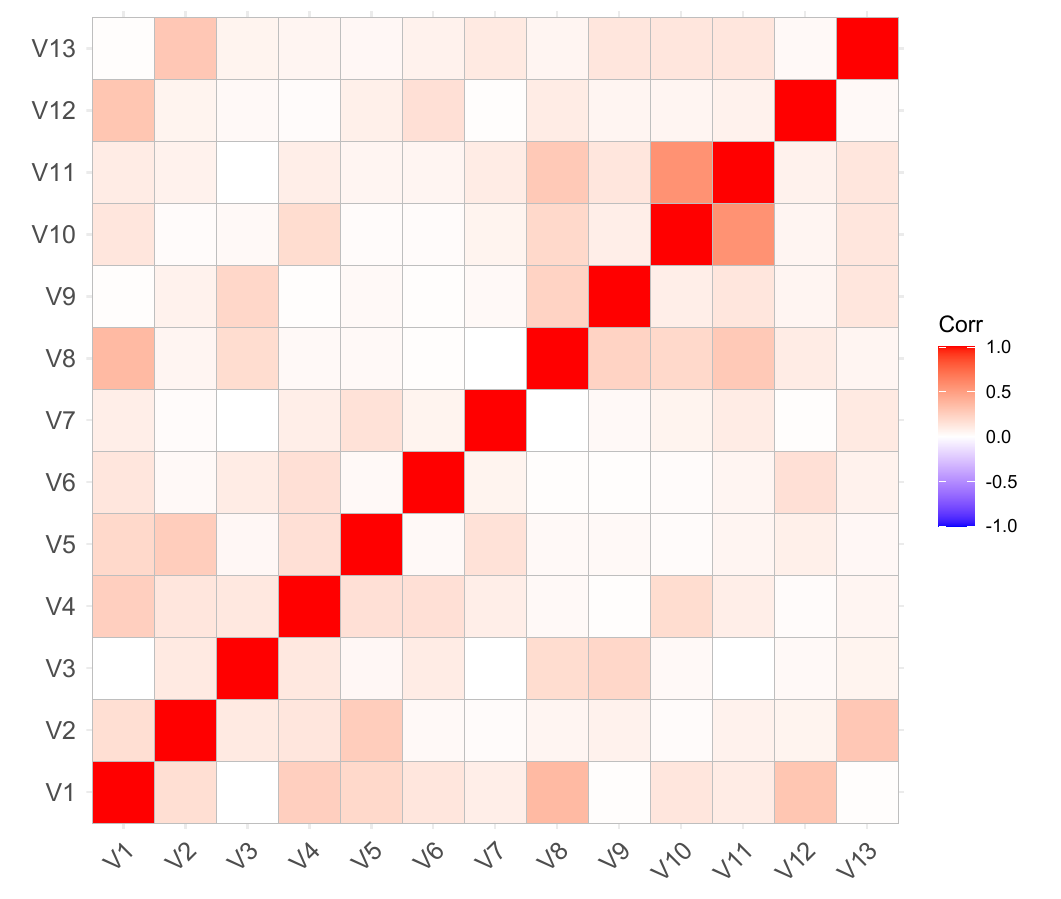}\hspace{0.01in}
\raisebox{-0.1in}{
\includegraphics[scale=0.36,clip, angle=0]{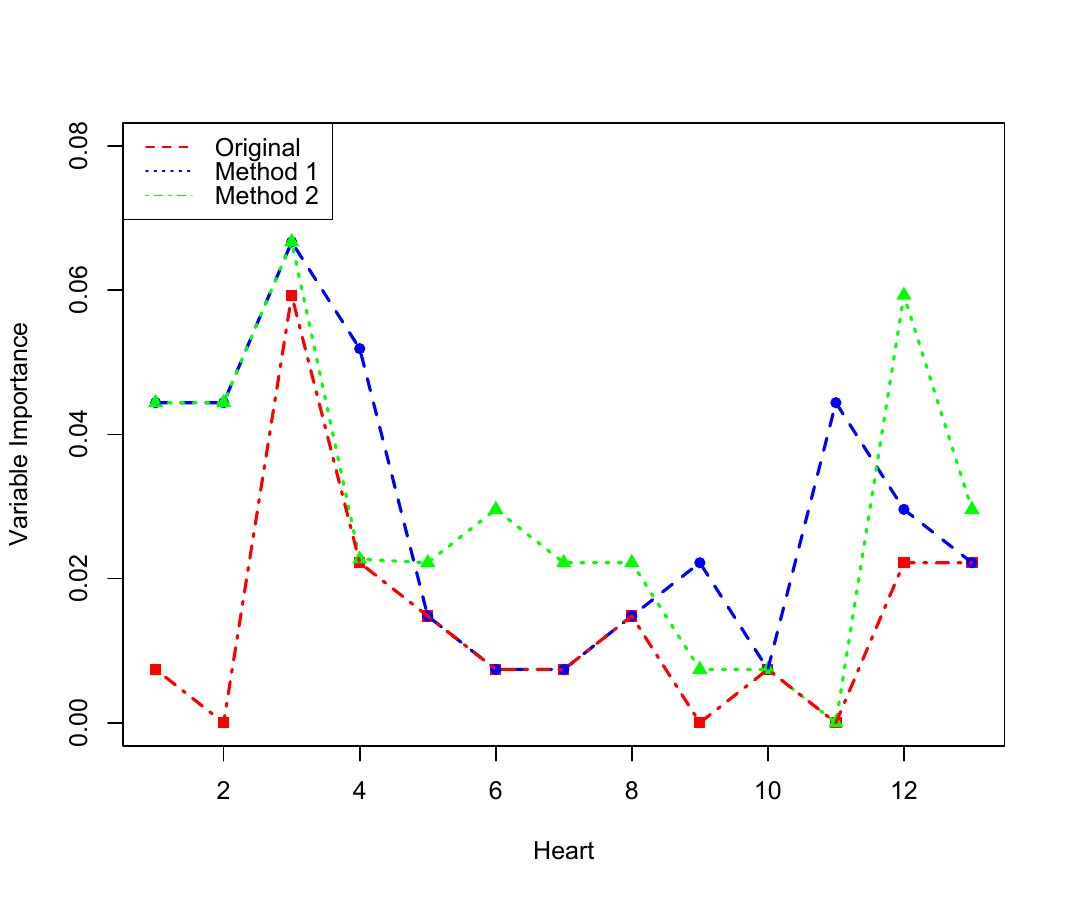}}
%%\abovecaptionskip=-1pt
\caption{\it The pairwise conditional correlation matrix of all variables in the UC Irvine Hearts dataset, and the importance
value given by the original RF, Method 1 and Method 2, respectively.}
\label{figure:gramHearts}
\end{center}
\end{figure} 
%%\\
\\
\\
For the UC Irvine Hearts dataset, Method 1 leads to the set of conditionally correlated variables for each
variable in the dataset as shown in Table~\ref{table:ccvHearts}. 
\begin{table}[htb]
\begin{center}
\begin{tabular}{r|l}
\hline
   \bf{Variable}                    				& \bf{Conditionally correlated variables}    \\
\hline 
$V_1$		&$V_8, V_{12}$\\
$V_2$		&$V_{5}, V_{13}$		\\
$V_3$			&$V_9$		\\
$V_4$			&$V_1, V_{10}$		\\
%%$V_5$			&		\\
$V_6$		&$V_4$\\
%%$V_7$		&		\\
$V_8$			&$V_3, V_{9}$		\\
$V_9$			&$V_3, V_{8}$		\\
$V_{10}$			&$V_{11}$		\\
$V_{11}$			&$V_8, V_{10}$		\\
$V_{12}$			&$V_{1}$		\\
\hline
\end{tabular}
\caption{\it Variables and their conditionally correlated variables for the UC Irvine Hearts dataset.} 
\label{table:ccvHearts}
%%\end{minipage}
\end{center}
%%\vskip -0.2in
\end{table}
The importance of two variables, age ($V_1$) 
and gender ($V_2$), are adjusted significantly upwards. Their importance scores, which are close to 0 by 
original RF, is clearly not reasonable. It is common knowledge that age plays a very important role in heart 
functions. Gender is critically important in heart studies \cite{PrabhavathiSelvi2014} because biological sex 
differences, e.g., hormones and smaller vessels, and gender-related factors such as symptoms, care disparities, 
and societal norms significantly impact heart disease presentation, diagnosis, treatment, and outcomes.
\\
\\
Resting blood pressure ($V_4$) is vital for heart health as it is a top indicator of cardiovascular risk. High 
resting blood pressure (hypertension) directly causes heart muscle thickening, scarring, and damage to 
artery walls, increasing risks of heart attacks, stroke, and heart failure. Accordingly, Method 1 assigns $V_4$
a very high importance score, which we think reasonable. In comparison, RF gives $V_4$ a moderate importance 
score, which is not sufficient to reflect its importance. 
\\
\\
Exercise-induced angina ($V_9$) is crucial in heart studies \cite{LindowEkstrom2021} as a strong indicator 
of underlying Coronary Artery Disease and a powerful predictor of future heart events like Acute Coronary Syndrome, 
revealing myocardial ischemia (lack of oxygen) during exertion, helping assess disease severity, and guiding 
treatment by showing how medications like nitroglycerin affect blood flow and demand, even revealing phenomena 
like ischemic preconditioning(the ``warm-up effect") that improve understanding of the heart's response to stress 
and potential regeneration. This supports the major upwards adjustment of $V_9$ by Method 1 from an importance 
score of nearly 0 as reported by the original RF.
\\
\\
The slope of the peak exercise ST segment ($V_{11}$) is a critical electrocardiographic marker \cite{FinkelhorNewhouse1986} 
for diagnosing coronary artery disease, with a slope indicating a positive test. It measures ST depression relative 
to heart rate changes, providing higher sensitivity than simple ST-segment depression alone. Thus $V_{11}$ is
also critical in heart studies or diagnosis. Accordingly, Method 1 corrects its nearly 0 importance, as reported by RF, 
to be of major importance.   
%%
%% 
%%\\
%%\\
%%\\
\\
\\
For Method 2, applying spectral clustering
with a proper choice of $\sigma$ leads to the following grouping of variables in the Hearts dataset:
~$\{V_1,V_8, V_{12}\}$, ~$\{V_2,V_5, V_{13}\}$, ~$\{V_3, V_9\}$, ~$\{V_4, V_6, V_7\}$, ~$\{V_{10}, V_{11}\}$. 
The resulting importance for most of the variables by Method 2 are similar to that by Method 1. The importance of two variables, age ($V_1$) 
and gender ($V_2$), are adjusted significantly upwards.
\\
\\
$V_6$ is Fasting Blood Sugar. Based on the recent research, there is a causal relationship between fasting blood glucose and myocardial infarction \cite{PengLiang2026}. 
However, RF and Method 1 only give a very low importance score the importance of $V_6$. 
$V_{12}$ is the number of major vessels colored by fluoroscopy. This metric measures blood vessels, which are the main vessels that supply blood to the heart. A higher number of vessels colored by fluoroscopy (0 to 3) is a diagnostic feature for heart disease \cite{Zhang2023}. Compare with RF and Method 1, Method 2 gives higher importance score which makes more sense. 
\\
\\
The other main difference between Method 1 and 2 is on
$V_4$ and $V_{11}$ which have no change of importance under Method 2. As argued above, $V_4$ appears to be one of the most important
variables, so correction by Method 1 makes sense while Method 2 is not reasonable in not adjusting the importance of $V_4$. 
Algorithmically, Method 2 fails to cover the correlation between $V_4$,  and $V_1$ thus the reported importance of $V_4$ is 
partially masked by $V_1$.  Same as $V_4$ , Method 2 fails to cover the correlation between  $V_{11}$ and $V_8$ thus the reported importance of $V_{11}$ is partially masked by $V_8$. 
\\
\\
The Hearts dataset originally contains 76 attributes, but all published experiments refer to using a subset of 14 of them with the Cleveland database in particular being the most commonly used database for studying heart disease patterns. So most of the variables should be important, and indeed variable importances reported by Method 1 and Method 2 (excluding $V_4$ and $V_{11}$) is consistent to this observation. 
\subsection{Wine dataset}
\label{data:Wine}
There are 13 variables in the UC Irvine Wine dataset, and are listed as follows: 1) Alcohol ($V_1$); ~2) Malic acid  ($V_2$); 
~3) Ash  ($V_3$); ~4) Alcalinity ($V_4$); ~5) Magnesium ($V_5$); 
~6) Total phenols  ($V_6$); ~7) Flavanoids ($V_7$); 
~8) Nonflavanoid phenols ($V_8$); ~9) Proanthocyanins  ($V_9$); 
~10) Color intensity ($V_{10}$); 
~11) Hue  ($V_{11}$); 
~12) OD280/OD315 of diluted wines  ($V_{12}$); 
~13) Proline ($V_{13}$).
The response variable is the quality (class label) of wine.
\begin{figure}[htb] 
\begin{center}
\includegraphics[scale=0.28,clip, angle=0]{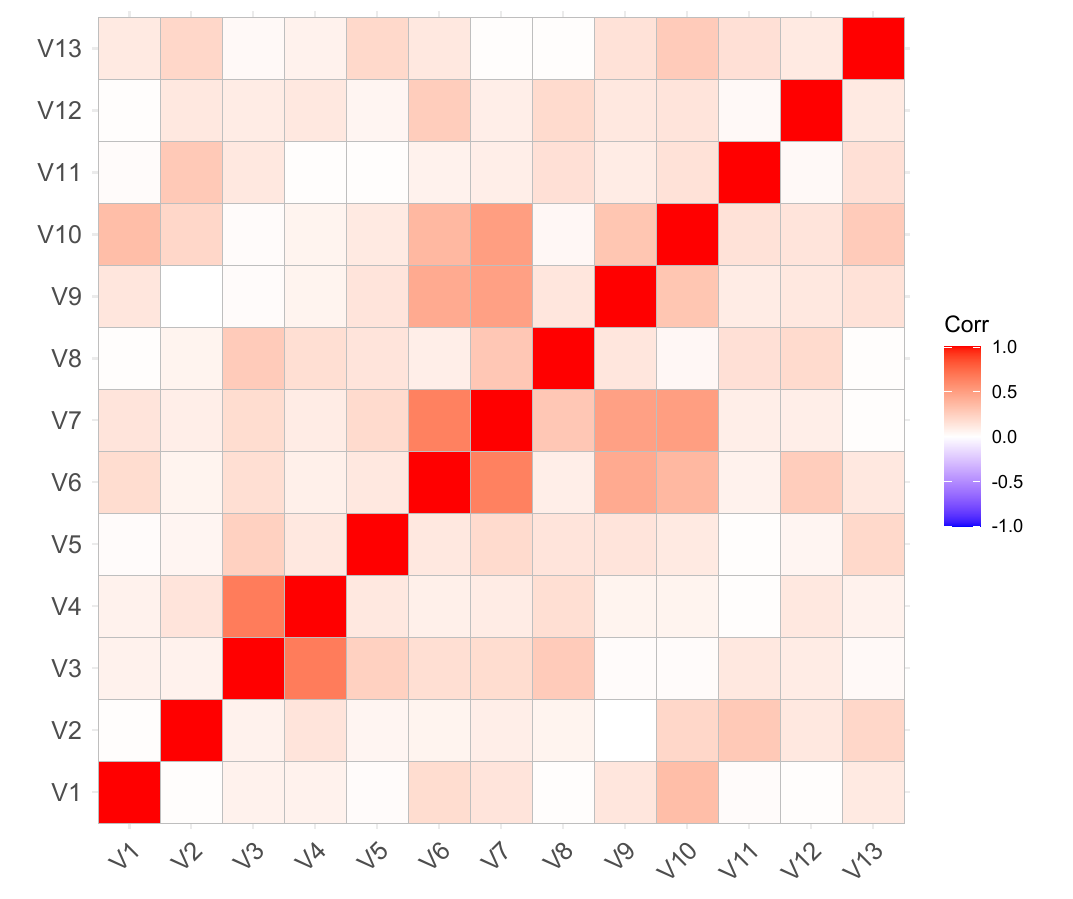}\hspace{0.01in}
\raisebox{-0.1in}{
\includegraphics[scale=0.35,clip, angle=0]{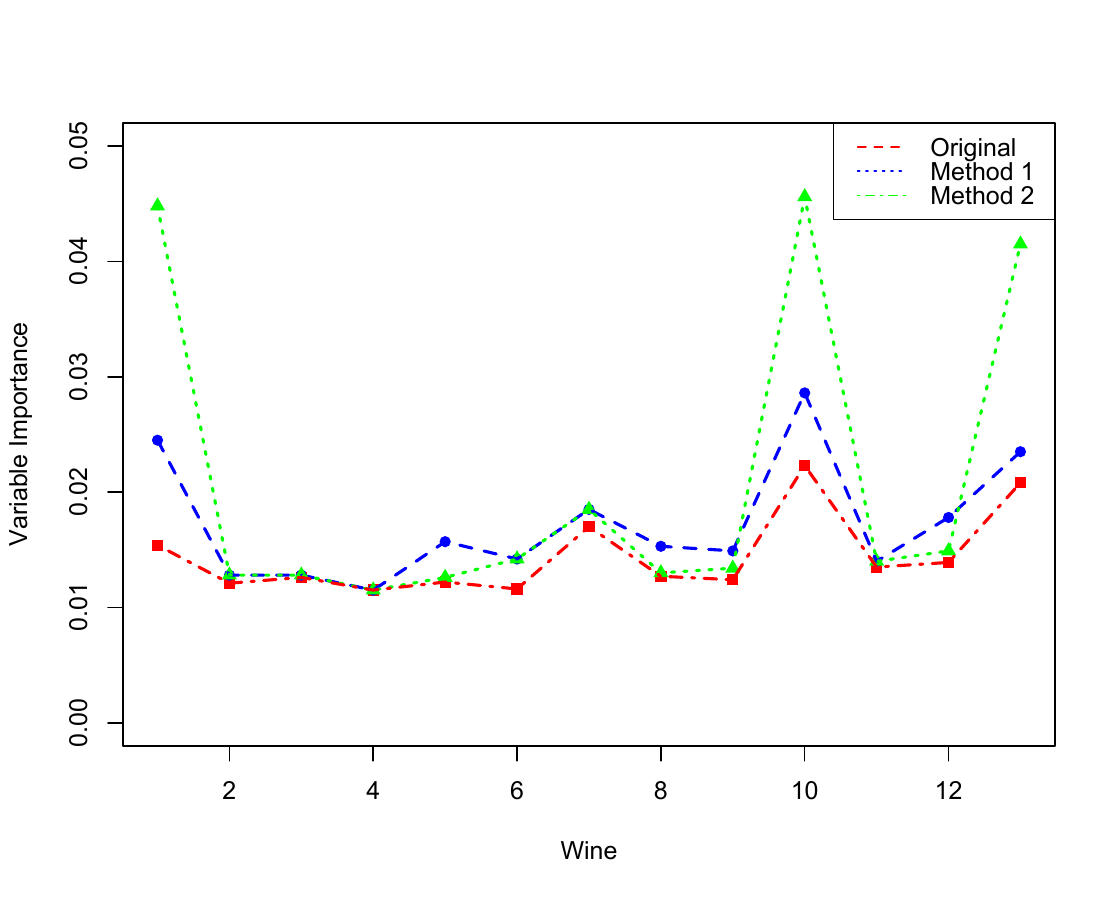}}
%%\abovecaptionskip=-1pt
\caption{\it The pairwise conditional correlation matrix of all variables in the UC Irvine Wine dataset, and the importance
value given by the original RF, Method 1 and Method 2, respectively.}
\label{figure:gramWine}
\end{center}
\end{figure} 
%%\\
\\
\\
For the UC Irvine Wine dataset, Method 1 leads to the set of conditionally correlated variables for each
variable in the dataset as shown in Table~\ref{table:ccvWine}.  
\begin{table}[H]
\begin{center}
\begin{tabular}{r|l}
\hline
   \bf{Variable}                    				& \bf{Conditionally correlated variables}    \\
\hline 
$V_1$			&$V_{10}$\\
$V_2$			&$V_{11}$		\\
$V_3, V_4$		&$V_3, V_4$		\\
$V_5$			&$V_3, V_7, V_{13}$			\\
$V_6, V_7$		&$V_6, V_7, V_9$		\\
$V_8$			&$V_3, V_7, V_{12}$			\\
$V_9$			&$V_6, V_7, V_{10}$			\\
$V_{10}$			&$V_7$			\\
$V_{11}$			&$V_2$			\\
$V_{12}$			&$V_6, V_8, V_{10}$			\\
$V_{13}$			&$V_2, V_5, V_{10}$			\\
\hline
\end{tabular}
\caption{\it Variables and their conditionally correlated variables for the UC Irvine Wine dataset.} 
\label{table:ccvWine}
%%\end{minipage}
\end{center}
%%\vskip -0.2in
\end{table}
\noindent
Alcohol ($V_1$) is a critical component of wine quality, acting as a structural, sensory, and preservative agent that shapes body, mouthfeel, 
and aroma. It contributes viscosity (body) to balance acidity and tannin, often creating a rounder, more ``generous'' texture in higher-alcohol 
wines. It acts as a carrier for aromas but can cause a ``hot'' sensation if excessive \cite{jordo_2015_from}. Accordingly, Method 1 assigns 
$V_1$ a very high importance score, which we think reasonable. In comparison, RF gives $V_1$ a moderate importance score, which is 
not sufficient to reflect its importance. 
\\
\\ 
For Method 2, applying spectral clustering with a proper choice of $\sigma$ leads to the following grouping of variables in the Wine dataset: 
~$\{V_1,V_{10},V_{13}\}$, ~$\{V_2, V_{11}\}$,  ~$\{V_3, V_4\}$, ~$\{V_5,V_8, V_{12}\}$,~$\{V_6,V_7, V_9\}$. The grouping results are very 
similar as in Method 1. The top 3 important variables are still  $V_1$,$V_{10}$ and $V_{13}$ in both Method 1 and 2. But Method 2 gives 
more significantly upward adjustment to $V_1$,$V_{10}$ and $V_{13}$. 
\subsection{Maternal health dataset}
\label{data:MaternalHealth}
There are 6 variables in the UC Irvine Maternal Health Risk dataset, and are listed as follows: 1) Age ($V_1$); ~2) Systolic 
Blood Pressure as SystolicBP ($V_2$); ~3)  Diastolic BP ($V_3$); ~4) Blood Sugar ($V_4$); ~5) Body Temperature as Body 
Temp ($V_5$); ~6) Heart Rate  ($V_6$). The response variable is the predicted risk intensity Level during pregnancy considering 
the previous attribute.
\begin{figure}[htb] 
\begin{center}
\includegraphics[scale=0.28,clip, angle=0]{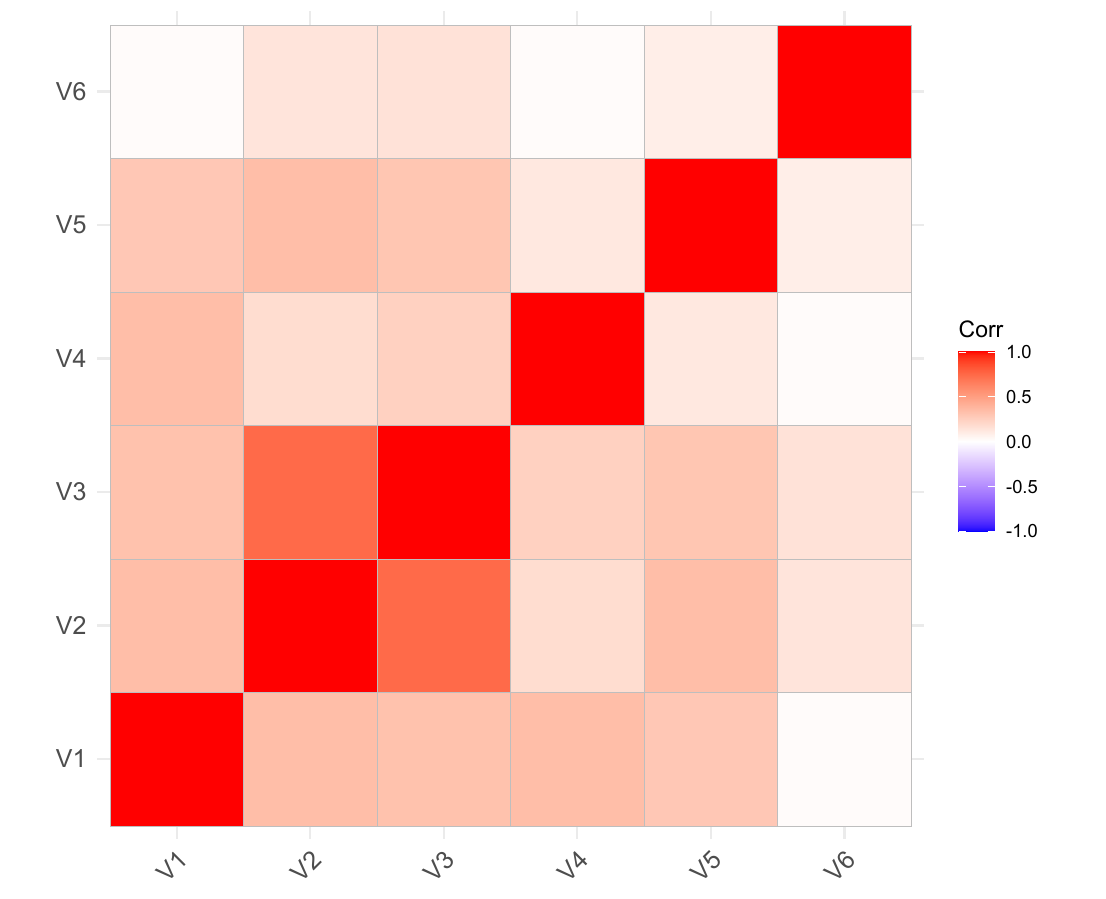}\hspace{0.01in}
\raisebox{-0.1in}{
\includegraphics[scale=0.36,clip, angle=0]{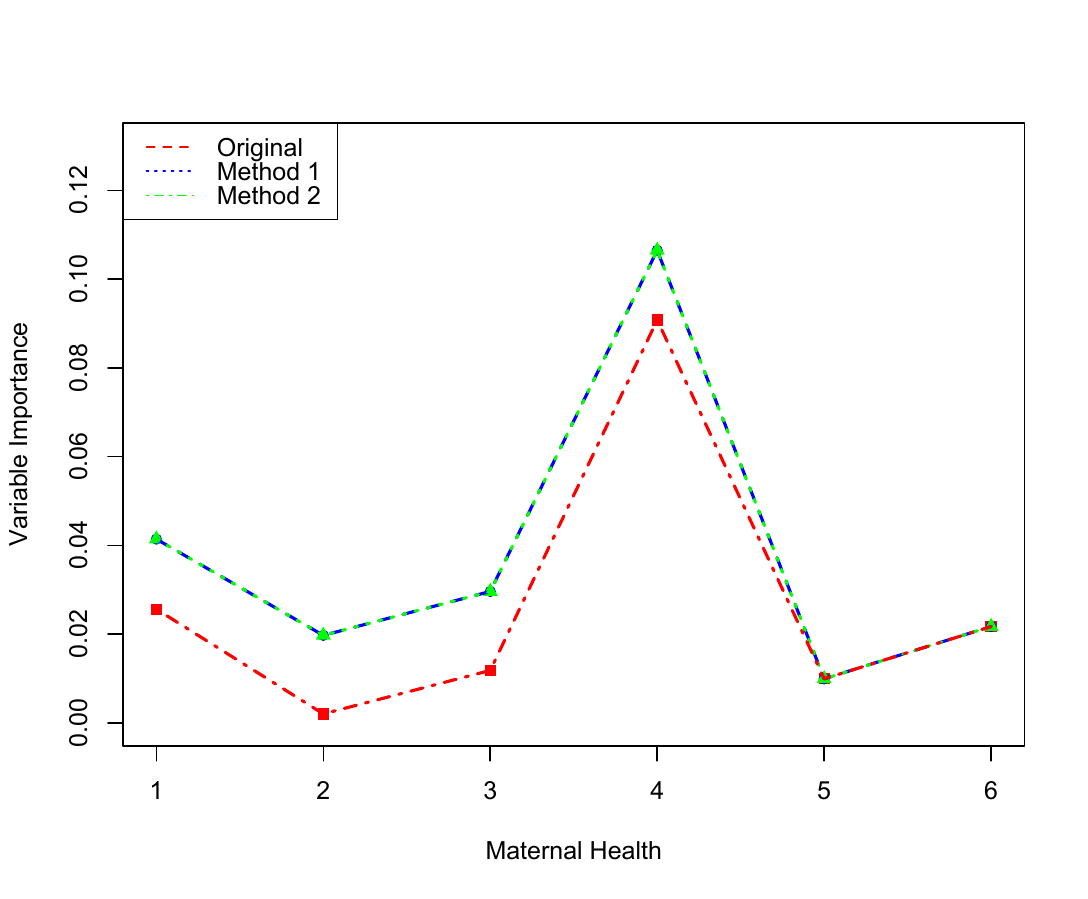}}
%%\abovecaptionskip=-1pt
\caption{\it The pairwise conditional correlation matrix of all variables in the UC Irvine Maternal Health dataset, and the importance
value given by the original RF, Method 1 and Method 2, respectively.}
\label{figure:gramMaternal}
\end{center}
\end{figure} 
%%\\
%%\\
For the UC Irvine Maternal Health datasets, Method 1 leads to the set conditionally correlated variables for each
variable in the dataset as shown in Table~\ref{table:ccvMaternal}.  
\begin{table}[H]
\begin{center}
\begin{tabular}{r|l}
\hline
   \bf{Variable}                    				& \bf{Conditionally correlated variables}    \\
\hline 
$V_1$		&$V_3$\\
$V_2$			&$V_3$		\\
$V_3$			&$V_1$		\\
$V_4$			&$V_3$		\\
%%$V_6$			&		\\
\hline
\end{tabular}
\caption{\it Variables and their conditionally correlated variables for the UC Maternal Health dataset.} 
\label{table:ccvMaternal}
%%\end{minipage}
\end{center}
%%\vskip -0.2in
\end{table}
\noindent
The two variables of SystolicBP (Upper value of Blood Pressure in mmHg) ($V_2$) and DiastolicBP (Lower value 
of Blood Pressure in mmHg) ($V_3$), are grouped together in Method 1. It is common knowledge that High Blood 
Pressure is considered with two factors: Upper ($V_2$)  and Lower ($V_3$) values of Blood Pressure. In 2017–2018, 
approximately 13 percent of women aged 18–39 years had chronic hypertension. Chronic hypertension has increased 
among pregnant women over time, largely due to increasing rates of obesity and increased maternal age. Women 
with hypertension are at higher risk for pregnancy complications such as superimposed preeclampsia, placental abruption 
(premature separation of the placenta from the uterus associated with abnormal bleeding), kidney failure, and cesarean 
delivery. Complications for the infant can also occur, including premature birth and fetal growth restriction. National 
guidelines now recommend that individuals self-monitor their blood pressure outside the clinical setting and work 
closely with their healthcare teams and others to achieve optimal control of their hypertension \cite{Osg2020}. 
Even in the original description of the dataset, it mentions Upper and Lower values of Blood Pressure in mmHg are 
significant attributes during pregnancy. In comparison, RF gives $V_2$ and $V_3$ moderate importance scores, 
which are not sufficient to reflect their importance.
\\
\\
For Method 2, applying spectral clustering with a proper choice of $\sigma$ leads to the following grouping of variables 
in the Maternal health dataset: ~$\{V_1,V_4\}$, ~$\{V_2,V_3\}$, ~$\{V_5\}$, ~$\{V_6\}$. It is exactly same as the grouping 
results by Method 1. 
So the resulting importance for all of the variables by Method 2 are the same to that by Method 1.
\subsection{Obesity dataset}
\label{data:Obesity}
There are 16 variables in the UC Irvine Obesity dataset, and are listed as follows: 1) Gender ($V_1$); ~2) Age ($V_2$); 
~3)  Height ($V_3$); ~4) Weight ($V_4$); ~5) Family history with overwieght ($V_5$); 
~6) FAVC: Do you eat high caloric food frequently? ($V_6$); ~7) FCVC: Do you usually eat vegetables in your meals? ($V_7$); 
~8) NCP: How many main meals do you have daily? ($V_8$); ~9) CAEC: Do you eat any food between meals? ($V_9$); 
~10) SMOKE: Do you smoke? ($V_{10}$); ~11) CH2O: How much water do you drink daily? ($V_{11}$); 
~12) SCC: Do you monitor the calories you eat daily? ($V_{12}$); ~13) FAF: How often do you have physical activity? ($V_{13}$); 
~14) TUE: How much time do you use technological devices such as cell phone, video games, television, computer and others? ($V_{14}$); 
~15) CALC: How often do you drink alcohol? ($V_{15}$); ~16) MTRANS: Which transportation do you usually use? ($V_{16}$); 
The response variable is the Obesity level.
\begin{figure}[htb] 
\begin{center}
\includegraphics[scale=0.28,clip, angle=0]{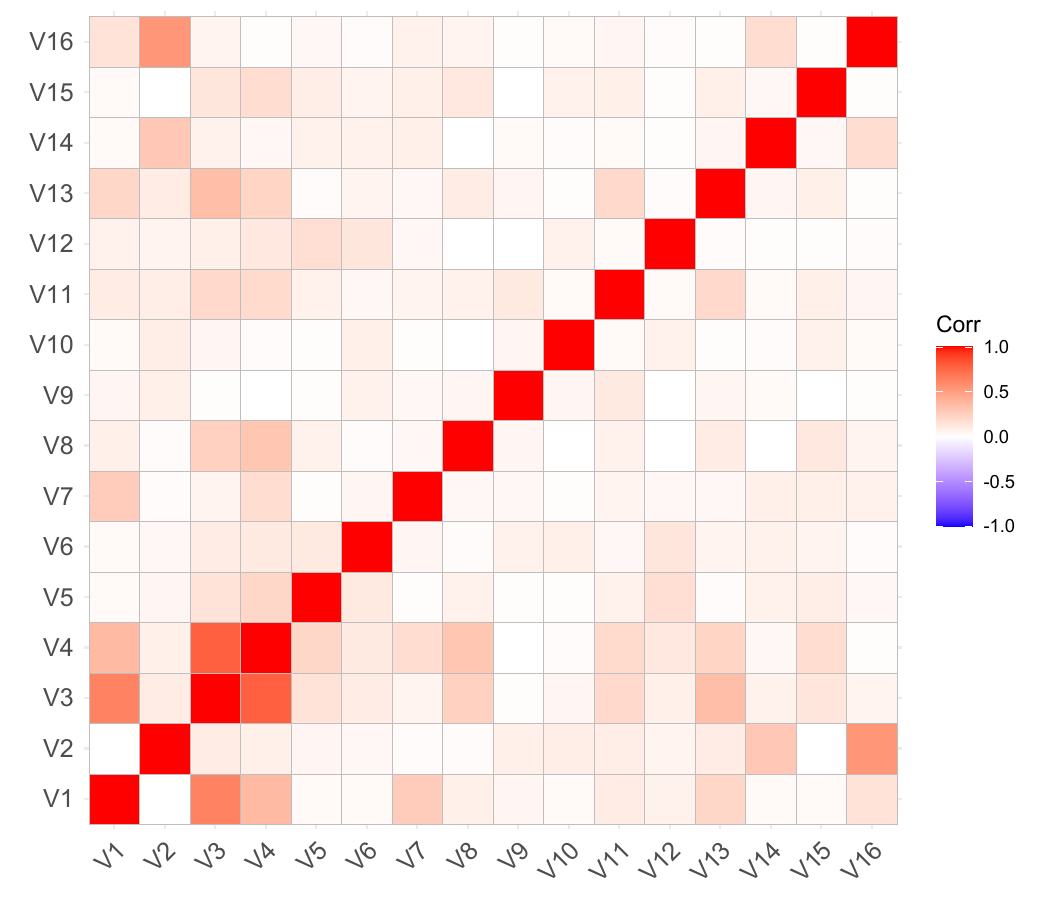}\hspace{0.01in}
\raisebox{-0.1in}{
\includegraphics[scale=0.34,clip, angle=0]{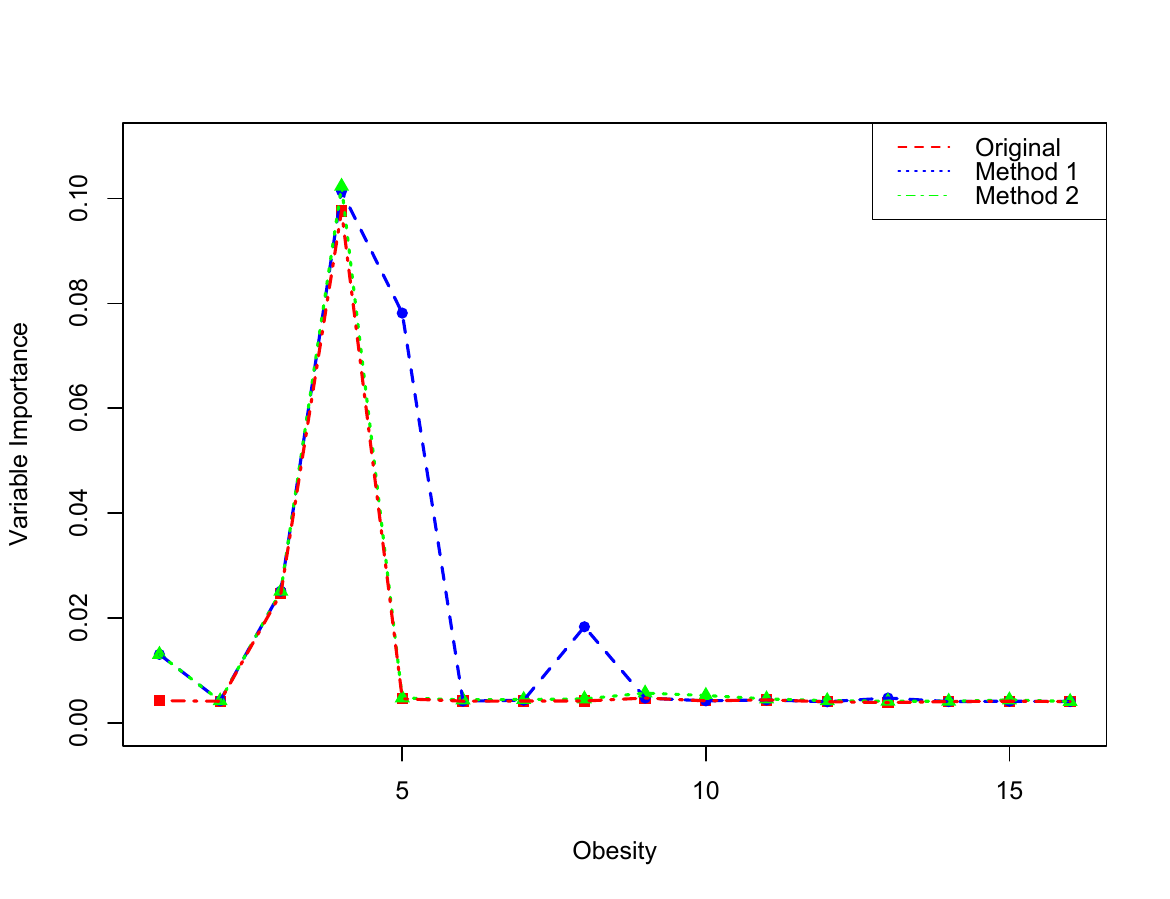}}
%%\abovecaptionskip=-1pt
\caption{\it The pairwise conditional correlation matrix of all variables in the UC Irvine Obesity dataset, and the importance
value given by the original RF, Method 1 and Method 2, respectively.}
\label{figure:gramObesity}
\end{center}
\end{figure} 
%%\\
%%\\
\\
\\
For the UC Irvine Obesity dataset, Method 1 leads to the set of conditionally correlated variables for each
variable in the dataset as shown in Table~\ref{table:ccvObesity}.  
\begin{table}[H]
\begin{center}
\begin{tabular}{r|l}
\hline
   \bf{Variable}                    				& \bf{Conditionally correlated variables}    \\
\hline 
$V_1, V_3, V_4$		&$V_1, V_3, V_4$\\
$V_2$			&$V_{14}, V_{16}$		\\
$V_5$			&$V_{4}, V_5, V_{12}$		\\
$V_7$			&$V_{1}$		\\
$V_8$			&$V_{3}, V_{4}$\\
$V_{13}$			&$V_{3}$\\
\hline
\end{tabular}
\caption{\it Variables and their conditionally correlated variables for the UC Irvine Obesity dataset.} 
\label{table:ccvObesity}
%%\end{minipage}
\end{center}
%%\vskip -0.2in
\end{table}
%%\\
%%\\
\noindent
Gender ($V_1$) differences significantly influence the prevalence of obesity, patterns of fat distribution, metabolic 
health outcomes, and responses to treatment. While women generally exhibit a higher overall prevalence of obesity, 
men are more susceptible to visceral fat accumulation, which increases the risk of cardiovascular disease (CVD), 
type 2 diabetes, and other obesity-related complications. This review examines the biological, genetic, and 
sociocultural foundations of sex-based differences in obesity \cite{KimKimSung2025}. Accordingly, Method 1 
corrects $V_1$, which has nearly 0 importance as reported by RF, to be of higher importance. 
\\
\\
For Method 2, applying spectral clustering with a proper choice of $\sigma$ leads to the following grouping of 
variables in the UC Irvine Obesity dataset: ~$\{V_1,V_3, V_4\}$, ~$\{V_2,V_{14}, V_{16}\}$, ~$\{V_5, V_6, V_{12}\}$, 
~$\{V_7, V_9, V_{10}\}$, ~$\{V_8,V_{11}, V_{13}, V_{15}\}$. 
The resulting importance for most of the variables by Method 2 are similar to that by Method 1. The importance of 
the variable Gender($V_1$), is adjusted properly. The only difference between Method 2 and 1 is the importance of  
$V_5$ which has importance of nearly 0 under Method 2.  Family history with overweight ($V_5$) is a crucial risk 
factor for obesity, with heritability estimates for body mass ranging from 55\% to 60\% in children. Familial predisposition, 
driven by both shared genetics and environmental habits, has the highest impact on children under 10. A child with two 
obese parents is significantly more likely to become obese, although not all children of obese parents will develop the 
condition \cite{NielsenNielsenHolm2015}. $V_5$ appears to be one of the most important variables, so correction by 
Method 1 makes sense while Method 2 is not reasonable in not adjusting the importance of $V_5$. Algorithmically, 
Method 2 fails to cover the correlation between $V_5$ and $V_4$ thus the reported importance of $V_5$ is partially 
masked by $V_4$.  
\subsection{Bank marketing dataset}
\label{data:BankMarketing}
There are 16 variables in the UC Irvine Bank Marketing dataset, and are listed as follows: 1) age ($V_1$); 
~2) job: type of job($V_2$); ~3)  marital: marital status ($V_3$); ~4) Education ($V_4$); ~5) default: has credit in default? ($V_5$); 
~6) balance: average yearly balance ($V_6$); ~7) housing: has housing loan? ($V_7$); ~8) loan:has personal loan? ($V_8$); 
~9) contact: contact communication type ($V_9$); ~10) day: last contact day of the month ($V_{10}$); 
~11) month: last contact month of year ($V_{11}$); ~12) duration: last contact duration ($V_{12}$); 
~13) campaign: number of contacts performed during this campaign and for this client  ($V_{13}$); 
~14) pdays: number of days that passed by after the client was last contacted from a previous campaign  ($V_{14}$); 
~15) previous: number of contacts performed before this campaign and for this client ($V_{15}$); 
~16) poutcome: outcome of the previous marketing ($V_{16}$); 
The response variable is the result that the client subscribed a term deposit or not.
\begin{figure}[htb] 
\begin{center}
%%\vspace{-0.05in}
\includegraphics[scale=0.28,clip, angle=0]{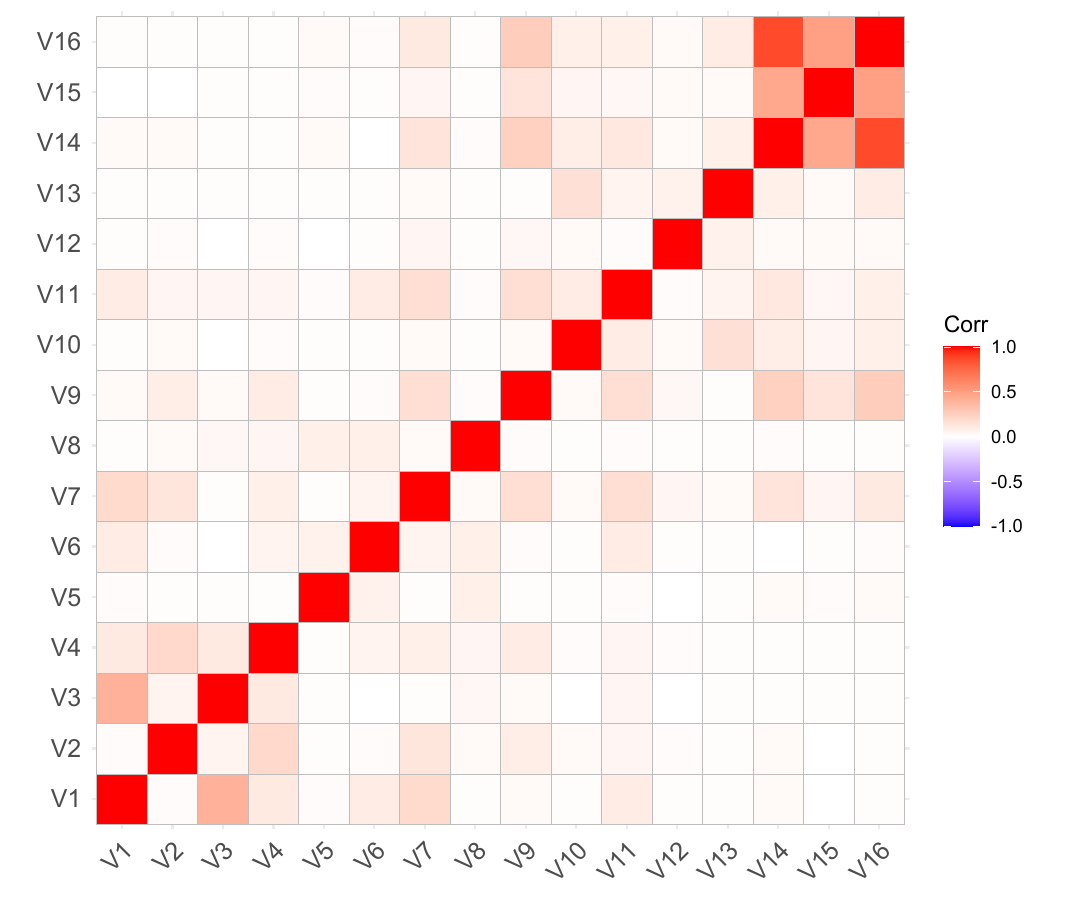}\hspace{0.01in}
\raisebox{-0.1in}{
\includegraphics[scale=0.36,clip, angle=0]{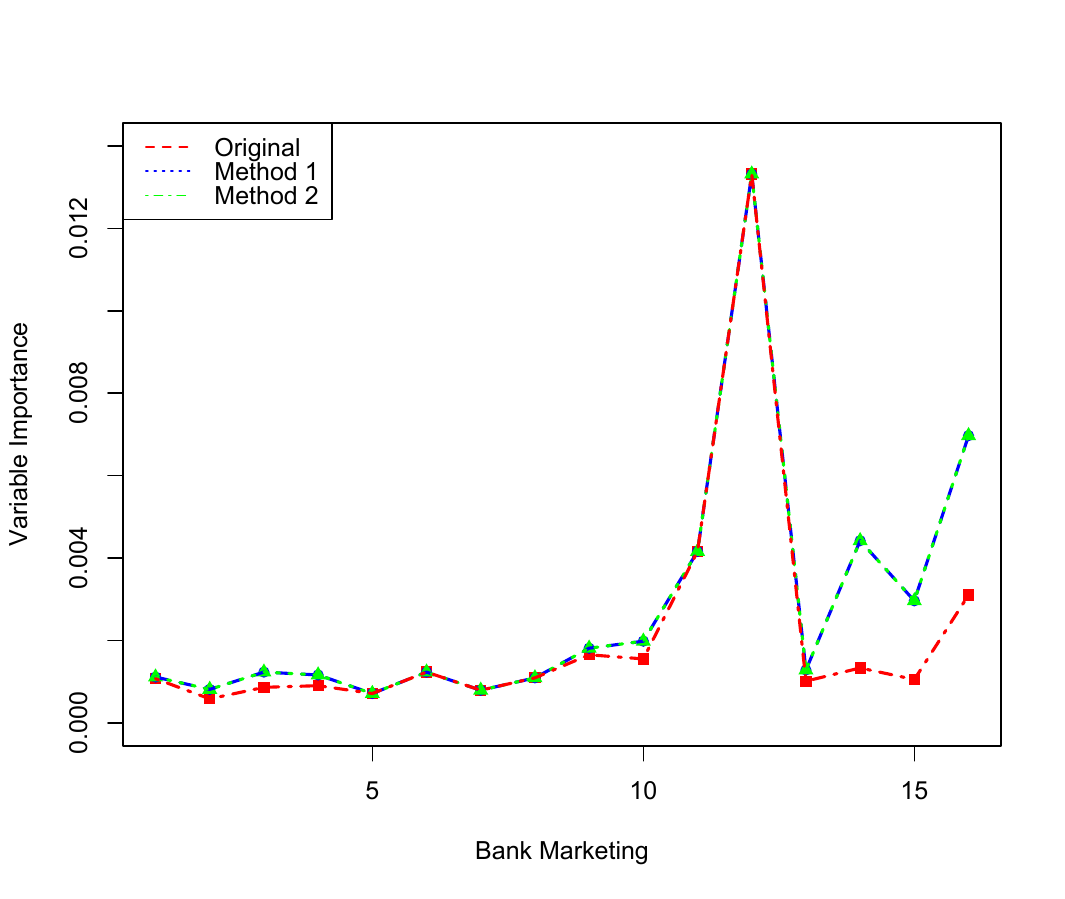}}
%%\abovecaptionskip=-1pt
\caption{\it The pairwise conditional correlation matrix of all variables in the UC Irvine Bank Marketing dataset, and the importance
value given by the original RF, Method 1 and Method 2, respectively.}
\label{figure:gramBank}
\end{center}
\end{figure} 
%%\\
%%\\
\\
\\
For the UC Irvine Bank marketing dataset, Method 1 leads to the set of conditionally correlated variables for each
variable in the dataset as shown in Table~\ref{table:ccvBank}.  
\begin{table}[htb]
\begin{center}
\begin{tabular}{r|l}
\hline
   \bf{Variable}                    				& \bf{Conditionally correlated variables}    \\
\hline 
$V_1, V_3$			&$V_1, V_3$\\
$V_2, V_4$			&$V_2, V_4$		\\
$V_9$			        &$V_{14}, V_{16}$		\\
$V_{10}, V_{13}$		&$V_{10}, V_{13}$		\\
$V_{14}, V_{15}, V_{16}$		&$V_{14}, V_{15}, V_{16}$		\\
\hline
\end{tabular}
\caption{\it Variables and their conditionally correlated variables for the UC Irvine Bank marketing dataset.} 
\label{table:ccvBank}
%%\end{minipage}
\end{center}
%%\vskip -0.2in
\end{table}
The importance of three variables, pdays: number of days that passed by after the client was last contacted from 
a previous campaign  ($V_{14}$), previous: number of contacts performed before this campaign and for this client 
($V_{15}$) and poutcome: outcome of the previous marketing ($V_{16}$) , are adjusted significantly upwards. 
($V_{14}$), ($V_{15}$) and ($V_{16}$)  can be considered as the Influence of Previous Contact History. Which 
makes sense that the customers who success in previous marketing campaign would have a high chance to 
subscribe a term deposit. The customers who are fail in previous marketing campaign would have a chance to 
not subscribe a term deposit \cite{LevyAppsflyer2025}. In comparison, RF gives very low importance scores, 
which are not sufficient to reflect their true importance. 
\\
\\
For Method 2, applying spectral clustering with a proper choice of $\sigma$ leads to the following grouping of 
variables in the UC Irvine Bank marketing dataset: ~$\{V_1,V_3\}$, ~$\{V_2,V_4\}$, ~$\{V_5\}$, ~$\{V_6\}$, 
~$\{V_7\}$, ~$\{V_8\}$, ~$\{V_9\}$, ~$\{V_{10}, V_{13}\}$, ~$\{V_{11}\}$, ~$\{V_{12}\}$, ~$\{V_{14}, V_{15}, V_{16}\}$. 
It is exactly same as the grouping results by Method 1. So the resulting importance for all of the variables by Method 2 
are the same to that by Method 1.
\section{Conclusions}
\label{section:conclusions}
This work attempts to correct the variable importance produced by RF. The idea of our approach is to remove influences from
correlated variables. We explore two effective options to achieve this, one by removing all correlated variables for
all individual variables, and the other by grouping variables according to their pairwise conditional correlation to other variables. 
Experiments on a number of UC Irvine benchmark datasets show that our approach achieves the expected corrections to variable
importance. While our work is primarily motivated by RF, it should be noted that it also applies to general predictive methods.  
Given the wide use of variable importance, we expect our approach be applied in many practical settings.
\section*{Acknowledgement}
The work of H. Xu and D. Yan are partially supported by Office of Naval Research (ONR) Grant, MUST IV, University of Massachusetts 
Dartmouth.
%%
%%
%%\bibliographystyle{plain}
%%\bibliographystyle{agsm}
%%\bibliography{../../myBib}

\end{document}